\documentclass{aa}  
\usepackage[draft]{todonotes}
\usepackage[colorlinks=true]{hyperref}
\hypersetup{linkcolor=blue, citecolor=blue}
\usepackage{graphicx}
\usepackage{txfonts}
\newcommand{\Enormresidual}{$\sigma_{r}$}
\newcommand{\Eresidual}{$E_{r}$}
\newcommand{\Emag}{$E_{mag}$}
\newcommand{\Ekin}{$E_{kin}$}

\newcommand{\MA}{$\mathcal{M}_{A}$}
\newcommand{\Ms}{$\mathcal{M}_{S}$}

\newcommand{\Prandtl}{Pr$_{\rm M}$}

\newcommand{\alfvenspeed}{$V_A$}
\newcommand{\alfvenratio}{$\mathcal{R_A}$}
\newcommand{\dbrms}{$\delta B_{rms}$}

\newcommand{\zplus}{$z^+$}
\newcommand{\zminus}{$z^-$}

\newcommand{\urms}{$u_{rms}$}
\newcommand{\teddy}{$t_{eddy}$}
\newcommand{\kforcing}{$k_f$}

\newcommand{\WL}{$W_L$}

\begin{document}

   \title{Residual energy in weakly compressible turbulence with a mean guide field}
  
   \titlerunning{Residual energy in magnetohydrodynamic turbulence}

   \subtitle{}

   \author{Raphael Skalidis\inst{1} \fnmsep \thanks{Hubble Fellow},
          Aris Tritsis\inst{2}, James R. Beattie\inst{3, 4}
          \and
          Philip F. Hopkins\inst{1}
          }

   \institute{TAPIR, Mailcode 350-17, California Institute of Technology, Pasadena, CA 91125, USA  \\
              \email{skalidis@caltech.edu} 
              \and
             Institute of Physics, Laboratory of Astrophysics, Ecole Polytechnique Fédérale de Lausanne (EPFL), Observatoire de Sauverny, 1290, Versoix, Switzerland 
             \and
    Department of Astrophysical Sciences, Princeton University, Princeton, 08540, NJ, USA
    \and
   Canadian Institute for Theoretical Astrophysics, University of Toronto, Toronto, M5S3H8, ON, Canada
             }

   \date{Received ; accepted}

 
  \abstract
   {Energy distribution is a fundamental property of magnetohydrodynamic (MHD) turbulence. In strongly magnetized turbulence, energy imbalances can arise, which can be quantified by the so-called residual energy: \Eresidual~=~(\Ekin~ - ~\Emag). Here, \Ekin\ and \Emag\ stand for the volume-averaged kinetic and magnetic energy, respectively. Numerical simulations of incompressible turbulence yield \Eresidual~$< 0$, which is consistent with solar wind observations; whereas in highly compressible turbulence simulations, we have \Eresidual~ $>$ 0. Differences arise in the cascade of \Eresidual\ between the two regimes.}
   {We explore the properties of \Eresidual\ in weakly compressible MHD turbulence in the presence of an initially strong (guide) magnetic field. We study the influence of different driving mechanisms and field strengths on the cascade of \Eresidual.}
   {We ran a suite of direct numerical simulations with the PENCIL code. All simulations were maintained through forcing in a quasi-static regime with sonic Mach numbers close to 0.1 for a number of different crossing times. We solely changed the Alfv\'en Mach number or, equivalently, the plasma beta ($\beta$) of the simulations. We drove turbulence by either injecting velocity or magnetic fluctuations at large scales and we studied the power spectra of kinetic, magnetic, density, and \Eresidual.}
   {Magnetically driven simulations show dominant Alfv\'enic fluctuations and a $\propto k^{-3/2}$ scaling in all spectra. In the inertial range \Eresidual\ $\approx$ 0. Turbulence is balanced with zero net cross-helicity, but exhibit strong locally imbalanced fluctuations that are qualitatively consistent with the dynamic alignment theory. Kinetically driven simulations give rise to a $\propto k^{-1}$ scaling, consistent with weakly interacting modes that preserve a high level of coherence throughout the inertial range. For the kinetically driven simulations, we report an excess in kinetic energy with respect to magnetic at all scales of the inertial range. The cascade of the residual energy (\Eresidual~$\propto k^\alpha$) depends on $\beta$ and scales as: for $\beta = 4.0$, $-2 \lesssim \alpha \lesssim -5/3$, for $\beta = 1.0$, $-5/3 \lesssim \alpha \lesssim -3/2$, and for $\beta = 0.3$, $\alpha \approx -1$.} 
{The energy partition in weakly compressible turbulence is strongly influenced by the forcing mechanism, even when the global sonic and Alfv\'enic Mach numbers are comparable across simulations.}


   \maketitle
%


\section{Introduction}
\label{sec:intro}

Magnetohydrodynamic (MHD) turbulence is a complex phenomenon occurring in many astrophysical systems. The energy distribution across spatial scales is a fundamental property of MHD turbulence that depends on key dimensionless parameters, such as the sonic (\Ms) and Alfv\'enic (\MA) Mach numbers, as well as the relevant driving mechanisms. 

Solar wind turbulence, which is weakly compressible and strongly magnetized, shows a disparity between kinetic and magnetic energy. This disparity is measured via the absolute (\Eresidual) or normalized (\Enormresidual) residual energy,
\begin{equation}
	\label{eq:residual_energy}
	E_r \equiv E_{kin} - E_{mag},  ~ \sigma_r \equiv \frac{E_{kin} - E_{mag}}{E_{kin} + E_{mag}}.
\end{equation}
In the solar wind \Eresidual~$< 0 $, indicating an excess of magnetic energy with respect to kinetic energy \citep{matthaeus_1982,tu_marsch_1995.book.waves.structures.solar.wind, perri_balogh_2010.crosshel.residual.Ulysses}. Additionally, the magnetic power spectrum is steeper than the kinetic. Similar results have been found in direct numerical simulations (DNS) of strongly magnetized and incompressible turbulence \citep{wang_2011.residual.energy.weak.turb.spotaneosuly.generated,shi_2025.residual.energy.intermittency,chen_2013.residual_energy.solar.wind}. 

In the linear regime, magnetic and velocity fluctuations are inseparable in individual Alfv\'en waves due to the symmetry of the incompressible MHD equations (i.e., \Eresidual~= 0). A net residual energy can only arise if nonlinear interactions of oppositely traveling Alfv\'en waves are considered \citep{boldyrev_2012.proceedings.residual.energy.perturbation.theory}. This phenomenon is observed in both weak and strong incompressible turbulence \citep{boldyrev_2012.proceedings.residual.energy.weak.strong.turb}. Other properties of incompressible turbulence can also lead to a net residual energy, such as cascade imbalances and symmetry breaks, imposed by the initial conditions \citep{dorfman_2025.residual.energy.symmetry.break.RMHD} or forcing schemes \citep{mason_boldyrev_2008.various.forcing.turb.spectrum,perez_2008.critical.balance.controled.by.forcing, lazarian_2025}. 

The resulting scaling law of the residual energy differs among different DNS setups. \cite{wang_2011.residual.energy.weak.turb.spotaneosuly.generated} simulated turbulence using the reduced MHD approximation \citep{tokamaks_1977} and found \Eresidual~$ \propto k_\perp^{-1}$. Others \citep{muller_grappin_2025.power_spectra.incompressible.MHD, boldyrev_2011.spectral_scalings.solarwind} have simulated incompressible MHD turbulence and derived \Eresidual~$ \propto k_\perp^{-2}$. Another approximation of incompressible turbulence, known as eddy damped quasi-normal Markovian (EDQNM), yields \Eresidual~$ \propto k_\perp^{-3/2}$ \citep{gogoberidze_2012.residual.energy.incompressible.EDQNM}. The various approximations to incompressible turbulence agree that nonlinear interactions are necessary to induce a net residual energy, but there are differences in the predicted spectral scalings, which can range from -1 to -2. 

Under incompressibility, the MHD equations can be written in a symmetric form using the Elsasser variable transformation \citep{elssaser_1950.variables}. In this regime, nonlinear interactions arise from oppositely traveling Alfv\'en waves, $\vec{z}^\pm$ (Sect.~\ref{sec:theoretical_consideration_magnetic}). Plasma inhomogeneities, however, break this symmetry. In compressible MHD, the properties of \zplus\ and \zminus\ are mixed even for low compressibility levels \citep{magyar_2019.elssaser.compressible.MHD}. Even in the weakly compressible regime, the generation of residual energy could be affected by a mixture of compressible and incompressible modes.

\cite{shi_2025.residual.energy.intermittency} derived that \Eresidual~$\propto k_\perp^{-2}$ in decaying and weakly (subsonic) compressible turbulence. This scaling is consistent with previous works of incompressible turbulence, but there are significant differences in the simulation setups. \cite{shi_2025.residual.energy.intermittency} simulated expanding boxes to account for the solar wind expansion and did not continuously drive turbulence. In contrast, the simulations of \cite{boldyrev_2011.spectral_scalings.solarwind} included continuous driving and a stationary box. 

In \cite{Vogel2011.thesis}, the properties of the residual energy were studied in highly compressible (supersonic), super-Alfv\'enic, and continuously forced turbulence. Contrary to Alfv\'enic turbulence simulations, \cite{Vogel2011.thesis} found a positive residual energy and a scaling that varies with the kinetic-to-magnetic energy ratio -- or equally the Alfv\'en ratio (\alfvenratio~$=$~\Ekin/\Emag) -- and the sonic Mach number. \cite{good_2025.pos.residual.energy.shocks} showed that fast-mode shocks can lead to positive residual energy. These results reveal major differences in the properties of \Eresidual\ between the incompressible and highly compressible regimes. 

The goal of this work is to link the incompressible and highly compressible regimes by investigating the properties of \Eresidual\ in weakly compressible turbulence. To achieve this, we ran DNS of sub-Alfv\'enic and weakly compressible forced turbulence. By explicitly evolving density perturbations, we sought to understand how weak density inhomogeneities might influence the cascade of \Eresidual. We solved the full set of single-fluid MHD equations, including dissipation and viscosity. In all runs, we maintained turbulence at a quasi-static regime with \Ms\ $\sim 0.1$ and considered three different initial magnetic field strength values: $B_0$ = 0.5, 1, and 2, corresponding to $\beta$ = 4.0, 1.0, and 0.3. We also explored the effects of driving (magnetic and kinetic) on the residual energy cascade. 

The organization of the paper is as follows: Sect.~\ref{sec:numerical_simulations} presents the numerical simulations. Sect.~\ref{sec:numerical_results} presents the main numerical results of the averaged and scale-dependent energetics. Sect.~\ref{sec:theory} addresses the fundamental differences between magnetically and kinetically driven turbulence based on the phenomenologies of dynamic alignment theory and wave dynamics influenced by inhomogeneities. Sect.~\ref{sec:discussion} discusses our obtained results in the context of solar wind turbulence. Sect.~\ref{sec:conclusions} summarizes the findings of this work.  

\begin{table*}[!htb]
\caption{Summary of numerical simulations} 
\label{table:sim_summary}
\centering                        
\begin{tabular}{c c c c c c c c c c c c c c} 
\hline\hline                
Resolution & Forcing & $\nu_3$ & $B_0$ \ & \Ms\ & \MA\ & $\beta$  & \dbrms\ &  \alfvenratio\  &  $E_c/E_s$    \\    
\hline                        
256$^3$ & K & $2.5\times 10^{-12}$ &0.5 & 0.08 & 0.16 & 4.0   & 0.05 & 2.39 $\pm$ 0.43  &  0.02  \\
256$^3$ & K & $2.5\times 10^{-12}$ & 1.0 & 0.08 & 0.08 & 1.0  & 0.05 & 2.55 $\pm$ 0.42  &  0.09  \\
256$^3$ & K & $2.5\times 10^{-12}$ & 2.0 & 0.09 & 0.05 & 0.3  & 0.05 & 3.43 $\pm$ 0.77  &  0.07  \\
512$^3$ & K & $8.0\times 10^{-14}$ & 2.0 & 0.09 & 0.05 & 0.3  & 0.04 & 3.50  $\pm$ 0.86 &  0.10 \\
256$^3$ & M & $2.5\times 10^{-12}$ & 0.5 & 0.08 & 0.16 & 4.0 & 0.09 & 0.70 $\pm$ 0.10  &  0.15 \\
256$^3$ & M & $2.5\times 10^{-12}$ & 1.0 & 0.09 & 0.09 & 1.0 & 0.09 & 0.75 $\pm$ 0.10  &  0.26  \\
256$^3$ & M & $2.5\times 10^{-12}$ &2.0 & 0.10 & 0.05 & 0.3  & 0.10 & 0.82 $\pm$ 0.10  &  0.23 \\
512$^3$ & M & $8.0\times 10^{-14}$ & 2.0 & 0.10 & 0.05 & 0.3 & 0.09 & 0.95 $\pm$ 0.10  &  0.38 \\
\hline
\hline
\end{tabular}
\tablefoot{K and M stand for kinetic and magnetic driving, respectively, while $E_c$ and $E_s$ correspond to the kinetic energy of compressible and solenoidal modes, respectively, as derived by the Helmholtz decomposition. The fiducial simulations have a resolution of $256^3$, whereas the $512^3$ simulations are included for convergence tests (Appendix~\ref{sec:numerical_convergence}).}
\end{table*}

\section{Numerical simulations}
\label{sec:numerical_simulations}

We solved the isothermal, hyper visco-resistive MHD equations with the PENCIL code \citep{pencil_collaboration_2021} via
\begin{equation}
\frac{D \vec{u}}{Dt} = - c_s^2 \nabla \ln{\rho} + \frac{1}{\rho} \vec{J} \times \vec{B} + F_{visc} + \vec{f_K}, 
\end{equation}
\begin{equation}
\frac{\partial \mathbf{B}}{\partial t} =  \nabla \times \left( \vec{u} \times \vec{B} \right) - (-1)^{n-1} \eta_n \nabla^{2n} \vec{B} + \vec{f_M},
\end{equation}
\begin{equation}
\frac{D \ln{\rho}}{Dt} = - \nabla \cdot \vec{u},
\end{equation}
where $\rho$, $\vec{u}$, $\vec{B}$, and $\vec{J}$ correspond to the gas density, velocity, magnetic field, and current density. The material derivative is $D = \partial/\partial t + \vec{u} \cdot \vec{\nabla}$. In all  simulations, we adopted periodic boundary conditions in a cubic domain of size $L$, so that all $g$ fields would satisfy $g(L\hat{e}_i)=g(0)$, where $\hat{e}_i$ denotes the unit vectors along the three Cartesian axes.

To maximize the inertial range in our simulations, we employed hyperviscous terms of third order. Hyper viscous force of order $n$  \citep{haugen_brandenburg2004.hyperviscosity} can be expressed as
\begin{equation}
    \label{eq:hyperviscous_force_general}
    F_{visc} = \frac {1}{\rho} \vec{\nabla} \cdot \left(2\rho \nu_{n} S^{(n)} \right ),
\end{equation}
where $\nu_{n}$ is the hyperviscosity and $S^{(n)}$ is the traceless rate of strain tensor; $n=1$ corresponds to normal viscosity. In all simulations, we considered $n=3$ and a constant $\nu_{3}$. The viscous force then becomes
\begin{equation}
    \label{eq:hyperviscous_force_n3}
    F_{visc} = \nu_{3}~\left( \nabla^6 \vec{u} + \frac{1}{3} \nabla^{4} \left[\vec{\nabla} \left( \vec{\nabla} \cdot \vec{u} \right) \right ] + 2S^{(3)} \cdot \vec{\nabla} \ln{\rho}
    \right) .
\end{equation}
The third term in the above equation does not appear in \cite{haugen_brandenburg2004.hyperviscosity}, who assumed that $\nu_{3}\rho = \rm const$. Similarly, in the induction equation we replaced normal diffusion with a constant hyperdiffusion $\eta_3$ following \cite{brandenburg_sarson_2002.hyperdiffusivity}. We chose both coefficients such that the Reynolds number at the grid scale is Re $\sim 3 - 5$, which is the stability limit of the code.

Hyperviscosity maximizes the inertial range of simulations by damping small-scale fluctuations. However, it can also lead to prominent bottleneck effects, by accumulating energy close to the dissipation scales, although the inertial range remains unaffected by hyperviscosity \citep{haugen_brandenburg2004.hyperviscosity}. On the other hand, hyperdiffusivity can affect large-scale dynamos \citep{brandenburg_sarson_2002.hyperdiffusivity}, which are suppressed in strongly magnetized turbulence, as in our simulations. Thus, we do not expect significant effects by the hyperviscous terms in the inertial range of our numerical setups.

We considered the following initial conditions, $\vec{B} = B_0 \vec{e_z}$, $\rho = 1$, $|\vec{u}| = 0$, and $c_s = 1$, expressed in dimensionless units. We considered the following $B_0$ magnitudes: 0.5, 1 and 2.0. We drove turbulence solenoidally at a quasi-statistically stationary regime, where the root mean square of the velocity (\urms) of all simulations is close to \urms~$\sim 0.1$, for approximately eleven eddy turnover times, where \teddy~$\equiv$~ $L$ / \urms. The driving function injects velocity and magnetic perturbations across a narrow bandwidth of $k$ modes ($k = 2\pi/\lambda$), which we set to \kforcing~$\sim 1.5$. We also tested forcing at \kforcing\ = 3, which produced results similar to \kforcing~$\sim 1.5$. Driving is performed by perturbing either the velocity (kinematic driving, $\vec{f_K}$) or the magnetic field (magnetic field, $\vec{f_M}$) \citep{brandenburg_oughton_2018.cross.helicity.decaying.turb} at a quasi-statistically stationary regime, where the root mean square of the velocity (\urms) of all simulations is close to \urms$\sim$ 0.1. 

Shear viscosity may impact the turbulence cascade of compressible and incompressible modes differently. This motivates the inclusion of a bulk viscosity ($\nu_{shock}$) to properly dissipate energy in compressible modes \citep{beattie_2025.compressibile.dynamo.bulk.visc}. The bulk viscous force is expressed as $F_{bulk} = \vec{\nabla} \left (\zeta_{\nu} \vec{\nabla} \cdot \vec{u} \right)$ with $\zeta_\nu = \nu_{shock}~ \langle \max(-\vec{\nabla}\cdot \vec{u}) \rangle~\delta x^2$, where $\nu_{shock}$ is a constant and $\delta x$ is the grid resolution \citep{gent_2013.pencil.SNe.multiphase,gent_2013.pencil.SN3.mean.field,gent_2020.SN.simulations.pencil}. In addition, $F_{bulk}$ is only applied in rarefactions, where discontinuities arise. In Appendix~\ref{sec:numerical_convergence}, we present numerical runs with different resolutions and $\nu_{shock}$. The strong agreement among the various numerical runs demonstrates the convergence of our results. The value of $\nu_{shock}$ has a negligible effect in the obtained spectra of our simulations, which are dominated by incompressible modes (Fig.~\ref{fig:power_spectra}). 

The regime of MHD turbulence is characterized by several key dimensionless quantities, which we define here for completeness. The sonic and Alfv\'en Mach numbers: $\mathcal{M}_S \equiv u_{rms}/{c_s}$, $\mathcal{M_A} \equiv u_{rms}/{V_A}$, where \alfvenspeed\ is the Alfv\'enic speed. The magnetic Prandtl number \Prandtl~$\equiv \eta~/\nu$ is equal to unity for all simulations. Plasma beta is defined as $\beta \equiv  \left ( \mathcal{M}_A / \mathcal{M}_S \right )^2$ and varies amongst runs. All simulations have \Ms~$\approx 0.1$. This choice allows us to focus on the weakly compressible regime and to isolate the effect of the driving mechanism and inhomogeneities on the turbulence energetics, while keeping the kinetic energy densities comparable across simulations. For the $256^3$ simulations, we used $\nu_3=2.5\times 10^{-12}$ which corresponds to an effective Reynolds number, Re $\sim u_{rms} \nu_3^{-1} k_f^{-5} \sim 5\times 10^{9}$. For the $512^3$ simulations, we considered $\nu_3=8\times 10^{-14}$, resulting in Re $\sim 1.5 \times 10^{11}$. At these resolutions, the viscosity and resistivity is therefore dominated by the numerical discretization \citep{Grehan2025_numerical_resistivity,Shivakumar2025_numerical_visc,Beattie2025_ISM_cascade_nature_astro}.  The simulation parameters along with additional details on the numerical resolution and/or the forcing mechanism are summarized in Table~\ref{table:sim_summary}.


\section{Numerical results}
\label{sec:numerical_results}

\begin{figure}
   \centering
   \includegraphics[width=\hsize]{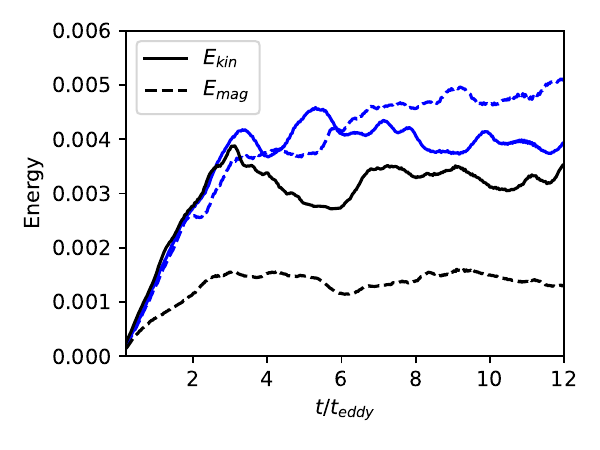}
      \caption{Time evolution of kinetic (solid) and magnetic (dashed) energy of kinetically (black) and magnetically (blue) driven turbulence with $\beta \approx 1$ and \Ms~$\approx 0.1$, corresponding to the second and sixth rows of Table~\ref{table:sim_summary}. Energy is in dimensionless units and time is normalized with the eddy turnover time. Driving affects the saturation level of the magnetic energy, yielding \alfvenratio~$\lesssim 1$ for magnetic, and \alfvenratio~$\approx 2.5$ for kinetic driving in the quasi-static regime, $t/t_{eddy}~\epsilon~[5, 12]$.}
         \label{fig:energy_time_profiles}
\end{figure}

\begin{figure}
   \centering
   \includegraphics[width=\hsize]{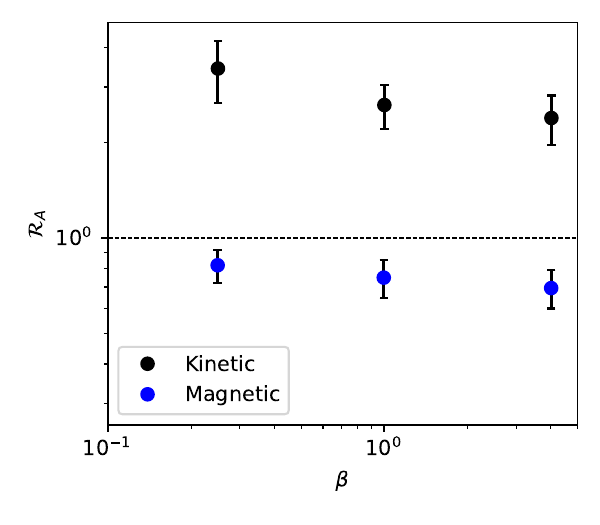}
   \caption{Alfv\'en ratio as a function of plasma beta of magnetically  (blue) and kinetically (black) driven turbulence. Horizontal line corresponds to perfect balance between kinetic and magnetic energies. The impact of forcing in the obtained volume-averaged energetics is evident: Kinetic driving yields \alfvenratio\ $> 1$ and magnetic driving \alfvenratio\ $\lesssim 1$,}
   \label{fig:ra_beta}
\end{figure}
    
\begin{figure}
   \centering
   \includegraphics[width=\hsize]{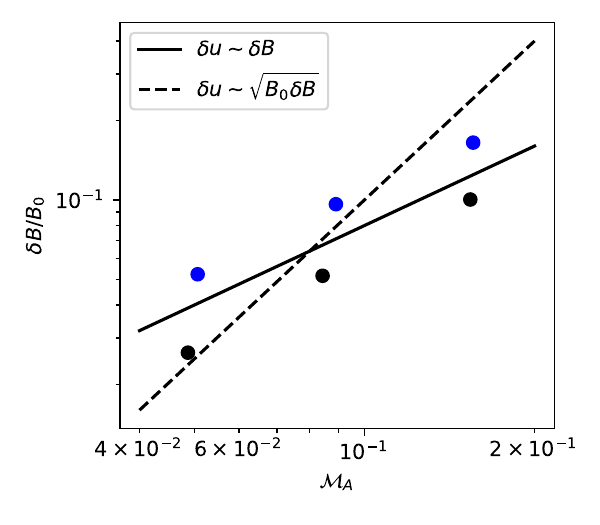}
    \caption{Fluctuating-to-order magnetic field ratio as a function of Alfv\'en Mach number. Kinetically and magnetically driven simulations are shown as black and blue dots, respectively. The solid black line corresponds to linear scaling $\delta u \sim \delta B$, while the dashed line corresponds to $\delta u \sim \sqrt{\delta B B_0}$. The numerical data strongly favor the linear scaling for both types of driving.}
    \label{fig:db_scaling}
\end{figure}

\begin{figure*}
   \centering
   \includegraphics[width=\hsize]{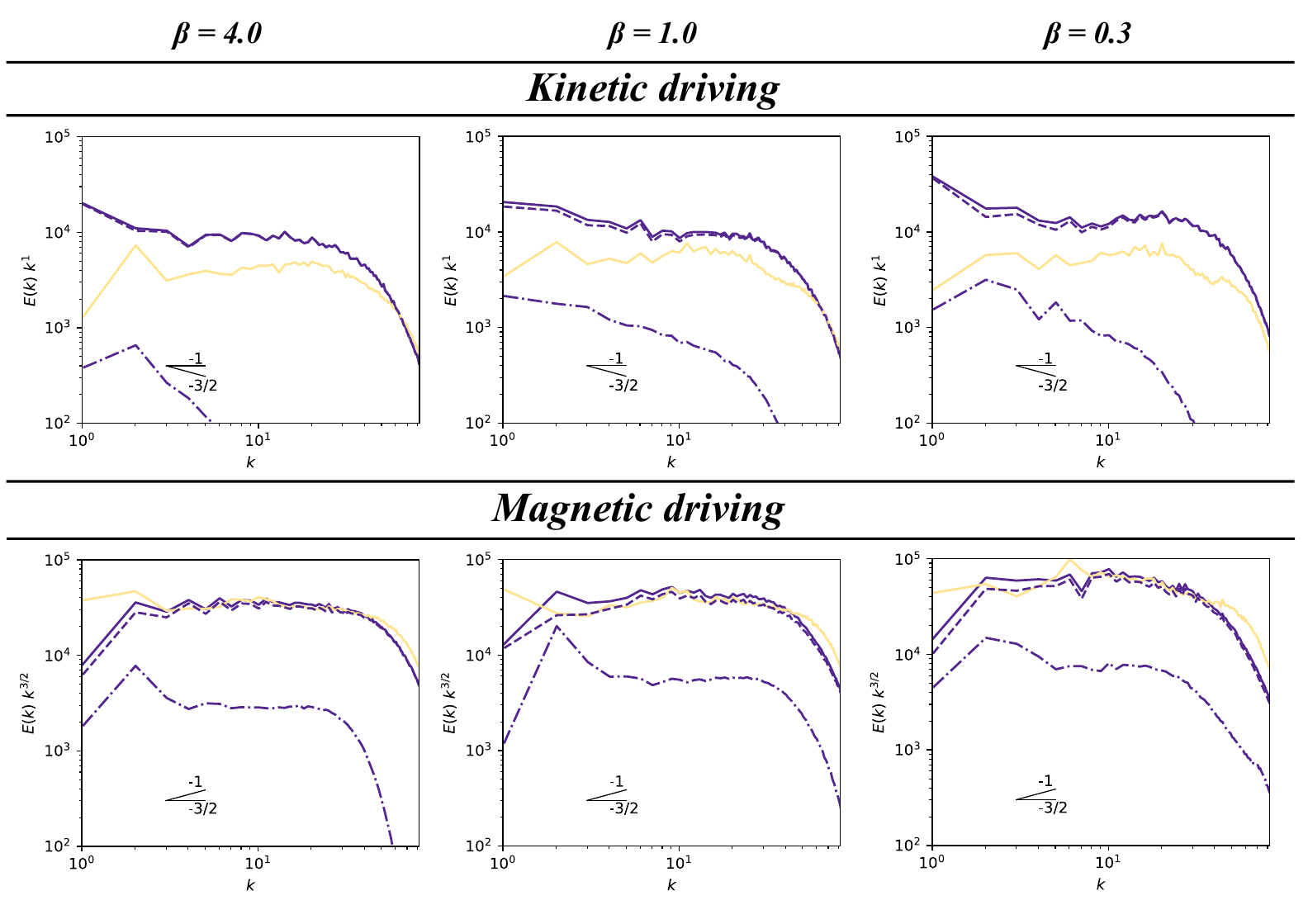}
    \caption{Kinetic (purple) and magnetic (yellow) compensated power spectra. Results of magnetically driven turbulence are shown in the bottom row, while of kinetically driven in the top. From left to right the initial magnetic field strength increases, or equivalently \MA\ (and $\beta$) decreases. Dashed and dashed-dotted black lines correspond to the kinetic power spectrum of the solenoidal and compressible modes, obtained from Helmholtz decomposition. The two power law scalings (-1 and -3/2) are shown for comparison. Incompressible modes carry the majority of kinetic energy in all simulations. In kinetically driven turbulence the turbulence cascade is shallower than magnetically driven simulations. There is a systematic excess in kinetic energy which leads to a positive residual energy. In kinetically driven simulations, the scaling of compressible modes is -3/2, which is different from the -1 scaling of incompressible modes.}
    \label{fig:power_spectra}
\end{figure*}

\begin{figure}
   \centering
   \includegraphics[width=\hsize]{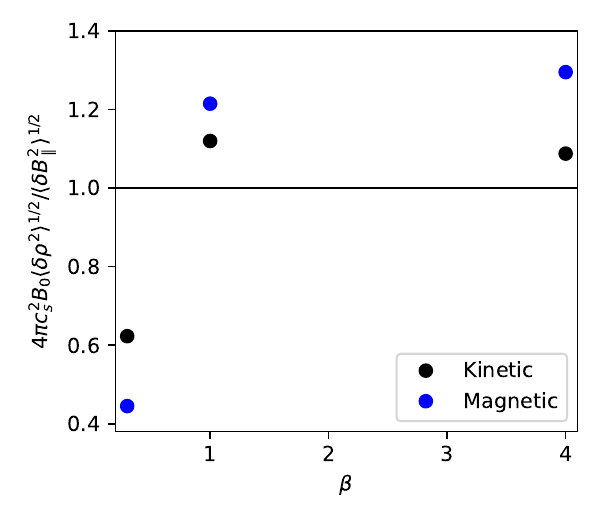}
   \caption{Slow mode pressure balance (Eq.~\ref{eq:slow_mode}) as a function of plasma beta. Black and blue points correspond to kinetic and magnetic driving, respectively. The slow mode relation accurately describes the numerical results, especially of simulations with $\beta \geq 1$.}
   \label{fig:slow_mode}
\end{figure}

\begin{figure*}
   \centering
   \includegraphics[width=\hsize]{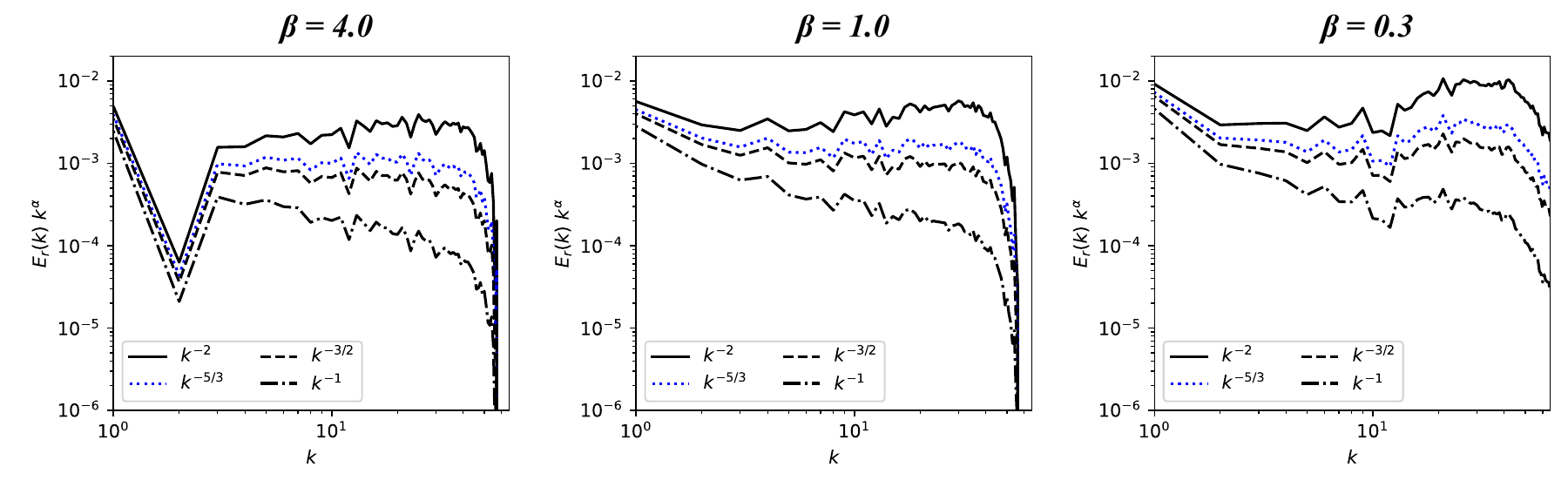}
    \caption{Compensated residual energy power spectra of kinetically driven turbulence for different $\beta$. Curves correspond to different power laws: $\propto k^{-2}$ (black solid), $\propto k^{-5/3}$ (blue dotted), $\propto k^{-3/2}$ (black dashed), and $\propto k^{-1}$ (black dash-dotted). For $\beta = 4.0$, we derive \Eresidual~$\propto k^{-3/2}$, for $\beta = 1.0$ that \Eresidual~$\propto k^{-3/2}$, while for $\beta = 0.3$ that \Eresidual~$\propto k^{-1}$. The scaling of the residual energy, which is always positive here, strongly depends on plasma beta.}
    \label{fig:er_cascade}
\end{figure*}   

All the results and figures presented below were extracted from a simulation snapshot at t=700 in code units or, equivalently, $t/t_{eddy} \approx 11$. Our fiducial simulations are those with a resolution of $256^3$ and $\nu_{shock}=0$. We have confirmed the robustness of these results using five additional snapshots in the quasi-static regime of turbulence, as well as higher resolution runs ($512^3$), which are not part of the main parameter study but are used to assess numerical convergence (Appendix~\ref{sec:numerical_convergence}).

\begin{figure*}
   \centering
   \includegraphics[width=\hsize]{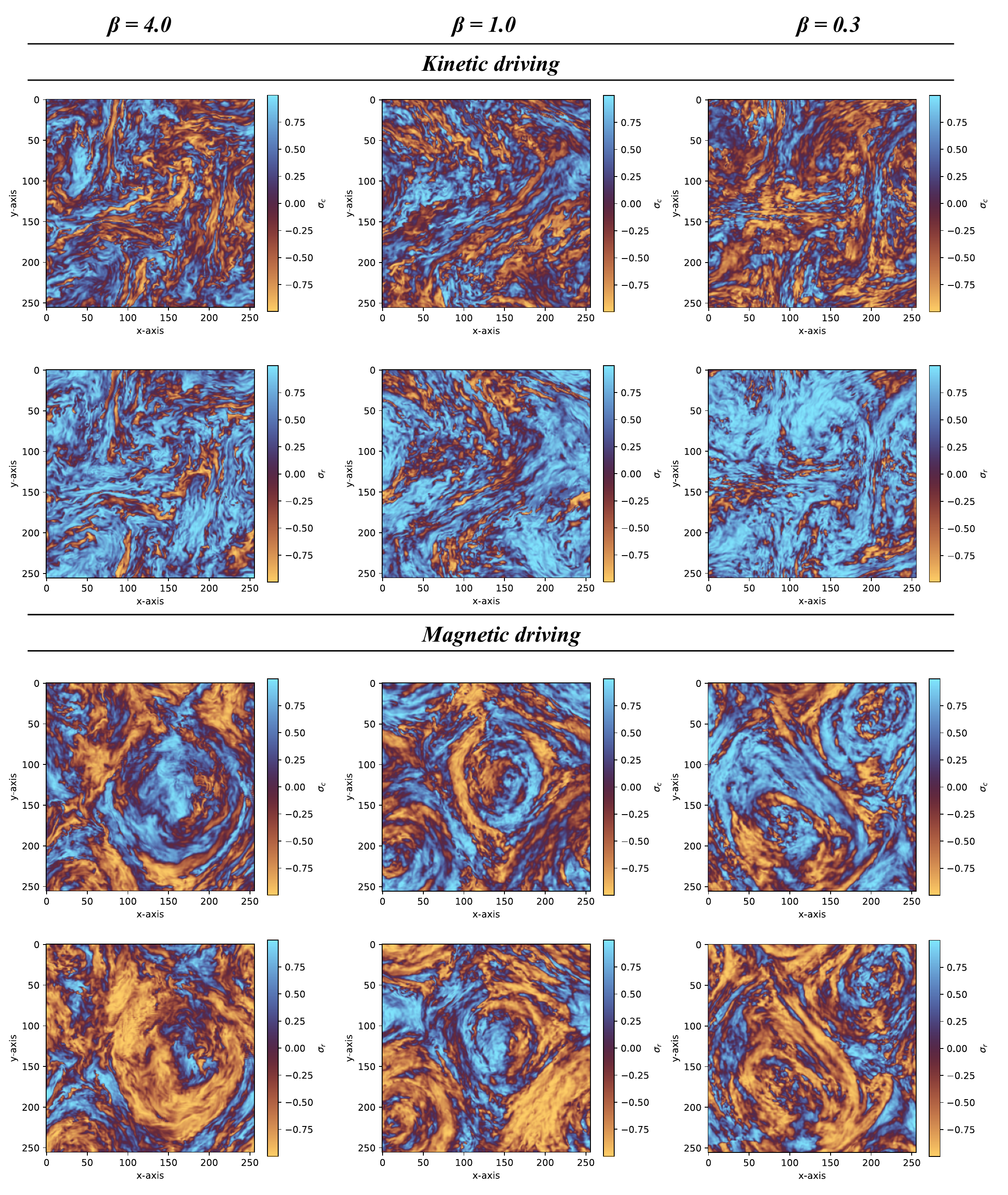}
    \caption{Normalized cross-helicity and residual energy 2D planes extracted from near the center of the simulation box. The magnetic field points towards the reader. The morphological structures between the depicted quantities are strongly correlated. Magnetically driven turbulence is dominated by local fluctuations of aligned and anti-aligned modes. These locally imbalanced regions lead to an overall balanced cascade, as predicted by the dynamical alignment theory. Kinetically driven turbulence shows more stochastic structure, weak alignment, and dominant velocity fluctuations ($\sigma_r \rightarrow 1$).}
    \label{fig:corss_hel_res_snapshots}
\end{figure*}

\subsection{Integral energetics}
\label{sec:averaged_energetics}

Fig.~\ref{fig:energy_time_profiles} shows the time evolution of the volume-averaged kinetic (solid curves) and magnetic (dashed curves) energies of two simulations with $\beta \approx 1$ each, driven magnetically (blue curves) and kinetically (black curves). Energies are shown in normalized units (dimensionless) and time is normalized over $t_{eddy}$.

Growth of the kinetic and magnetic energies occurs for $t/t_{eddy} \lesssim 3$, while for $t/t_{eddy} \gtrsim 3$ kinetic energy reaches the quasi-static regime. As is evident from Fig.~\ref{fig:energy_time_profiles}, the growth rate of \Emag\ and its saturation level depends on forcing, with magnetic forcing leading to more efficient growth. We found similar results in all simulations. For kinetically driven simulations, we also found that the timescale of the exponential growth phase of \Emag\ depends on $|\vec{B}_0|$ with lower values yielding a faster saturation (not shown here). This happens because dynamo is suppressed when strong magnetic fields are present \citep{haugen_brandenburg_2004.large.scale.dynamo.supression}. The kinetic energy densities of all simulations are comparable, as we have targeted for approximately the same \Ms. Energy differences arise because of the various saturation levels in \Emag. 

Fig.~\ref{fig:ra_beta} depicts \alfvenratio\ as a function of plasma beta. Black points correspond to kinetically driven turbulence, while blue points to magnetically driven. Error bars indicate the standard deviation of \alfvenratio\ measured over time. For magnetically driven simulations \alfvenratio\ ranges from 0.7 to 0.8, whereas for kinetically driven simulations it ranges from 2.5 to 3.4. In magnetically driven turbulence, the magnetic excess is due to fluctuations at the injection scale, while in kinetically driven turbulence the kinetic excess appears at all scales (Sects.~\ref{sec:scaling_magnetic_driving} and~\ref{sec:scaling_kinetic_driving}). In the kinetically driven simulations, the marginal increment of the \alfvenratio\ with $B_0$, corresponding to lower \MA, is due to the slightly higher \Ms\ of these simulations (Table~\ref{table:sim_summary}). However, given the obtained standard deviations, these differences are considered insignificant. 

To understand the energy ratio differences, we consider the time evolution equation of the volume-averaged kinetic and magnetic energies, considering triply periodic boundary conditions \citep{haugen_brandenburg_2004.large.scale.dynamo.supression, brandenburg_2014.energy.ratio.prandtl}:
\begin{align}
    \label{eq:energy_time_evolution}
    & \left\langle \frac{d E_{kin}}{dt} \right\rangle   =    + \langle \vec{u} \cdot \left ( \vec{J} \times \vec{B} \right) \rangle  - Q_\nu + W_{f_K}  + c_s^{2} \langle \rho \vec{\nabla} \cdot \vec{u} \rangle \\
    \label{eq:energy_time_evolution_magnetic}
    & \left\langle \frac{d E_{mag}}{dt} \right\rangle            =    - \langle \vec{u} \cdot \left ( \vec{J} \times \vec{B} \right) \rangle   - Q_\eta + W_{f_M}.
\end{align}

We note that the magnetic evolution equation (Eq.~\ref{eq:energy_time_evolution_magnetic}) is derived from the induction equation and includes the term $\langle \nabla \cdot \left[ \vec{B} \times (\vec{u} \times \vec{B}) \right] \rangle$, which vanishes by virtue of the divergence theorem, $\int_V dV ~ \vec{\nabla} \cdot \left[ \vec{B} \times \left(\vec{u} \times \vec{B} \right) \right] = \oint_S dS~  \vec{n} \cdot \left[ \vec{B} \times \left(\vec{u} \times \vec{B} \right) \right] = 0$, where $\vec{n}$ denotes the outward normal vector on the surface $\vec{S}$. For periodic boundary conditions, the values of all fields are identical on opposite sides of the simulation domain, so that the flux through opposing edges of the simulations cancel exactly, yielding a zero net surface integral.

The first term on the right hand side in Eqs.~\ref{eq:energy_time_evolution} and \ref{eq:energy_time_evolution_magnetic} corresponds to the work exerted by the Lorentz force (\WL). Viscous dissipation and Joule heating energy rates are denoted as $Q_\nu$ and $Q_\eta$, respectively. The energy rates of kinetic and magnetic driving correspond to $W_{f_K}$ and $W_{f_M}$. 

The sign of the Lorenz force determines the energy exchange rate between \Ekin\ and \Emag. For \WL~$< 0$, kinetic energy transfers to magnetic, while for \WL~$> 0$, the magnetic energy transfers to kinetic. In the quasi-static regime, the left-hand sides in the above equations are zero. We can then derive
\begin{equation}
	\label{eq:energy_equipartition}
	W_f + W_P \sim Q_\nu + Q_\eta,
\end{equation}
where $W_P= c_s^{2} \langle \rho \vec{\nabla} \cdot \vec{u} \rangle$ and $W_f = W_{f_K} + W_{f_M}$, which is the total energy rate due to forcing, including both magnetic and velocity components. 

In incompressible turbulence, $W_P = 0$, whereas in compressible turbulence, we have $W_P \neq 0$. The relative contribution of $W_P$ to the energy balance (Eq.~\ref{eq:energy_equipartition}) depends on the correlation between $\rho$ and $\vec{\nabla} \cdot \vec{u}$. In weakly compressible turbulence, the majority of modes are solenoidal (Sect.~\ref{sec:scaling_properties}); hence, we have $W_f \gg W_P$. Thus, we find that $W_f \sim Q_\nu + Q_\eta$, which is similar to leading order to incompressible turbulence. Following \cite{wei_2025.alven.ratio.length.scale.dependence}, we assumed an isotropic, constant dissipation rate, and $Q_\nu \sim Q_\eta$. For \Prandtl~= 1, we obtained \alfvenratio~$\sim \left( \ell_u/\ell_b \right)^2$, where $\ell_u$ and $\ell_b$ are the correlation scales of velocity and magnetic fluctuations, respectively.

The correlation length is defined as \citep[e.g.,][]{beattie_2025.compressibile.dynamo.bulk.visc,connor_2025.SN.turbulence}:
\begin{equation}
	\ell_g \equiv \frac{2\pi}{L} \frac{\int dk\, E(k) k^{-1}}{\int dk\, E(k)},
\end{equation}
where $E(k)$ is the power spectrum of a field $g$ (magnetic field, velocity, or density). In this work, we calculated the shell-integrated power spectrum properties via
\begin{equation}
  \label{eq:power_spectrum}
  E(k) = \frac{1}{N} \int d\vec{\Omega}_k ~ |\hat{g}(\vec{k})|^2~4\pi k^2,
\end{equation}
where $\hat{g}$ is the Fourier transform of a field $g$ (e.g., velocity), and $\Omega_k$ is the spherical shell with radius $1/k$ within which integration takes place. The normalization factor $N$ is calculated from Parseval's theorem. 
 
Through the kinetically driven simulations, we derived $\ell_u \sim 0.5$ and $\ell_b \sim 0.25$, which yields $\ell_u^2 / \ell_b^2 \sim 4$. In the magnetically driven simulations, we obtained $\ell_u \sim 0.4$ and $\ell_b \sim 0.5$; hence, $\ell_u^2 / \ell_b^2 \sim 0.6$. Thus, the differences in the correlation lengths influence the energy ratios, as predicted by \cite{wei_2025.alven.ratio.length.scale.dependence}. Because forcing affects the correlation lengths, it has a direct impact on the obtained energy ratios. 

Forced fields have larger correlation lengths and approximate the driving scale, which is  $\ell_f \sim 0.5$. For this reason, in kinetically driven turbulence, we find that $\ell_u \sim 0.5$ , while in magnetically driven, this is $\ell_b \sim 0.5$. The correlation lengths of unforced fields develop naturally from nonlinear interactions. 

Energy conversion from kinetic to magnetic (kinetic forcing) happens via small-scale dynamo because our simulations are sub-Alfv\'enic. The efficiency of small-scale dynamo depends on a number of parameters, like \MA, \Ms, and viscosity, which could influence $\ell_b$ and, hence, \alfvenratio\ \citep{haugen_brandenburg_2004.large.scale.dynamo.supression, beattie_2025.compressibile.dynamo.bulk.visc}. 

This leads us to the first conclusion about the volume-averaged energetics: \alfvenratio\ > 1 for kinetic driving, and \alfvenratio\ $\leq$ 1 for magnetic driving, corresponding to \Eresidual\ > 0 and \Eresidual\ < 0 for kinetic and magnetic driving, respectively. Our conclusion is broadly consistent with recent simulations of incompressible turbulence \citep{lazarian_2025}, although differences arise regarding the relative scaling between magnetic and kinetic fluctuations. 

\cite{lazarian_2025} found that $\delta u \sim \sqrt{B_0 \delta B}$ for kinetic driving, and $\delta u \sim \delta B$ for magnetic driving in incompressible turbulence. However, our simulations of weakly compressible turbulence strongly suggest an Alfv\'enic scaling ($\delta u \sim \delta B$) for both types of driving. Fig.~\ref{fig:db_scaling} shows the fluctuating-to-order magnetic field ratio as a function of the Alfv\'en Mach number. Black points correspond to time- and volume-averaged properties of kinetically driven, while blue points to magnetically driven turbulence. Numerical points are consistent with a linear scaling, instead of a square root scaling as shown by the black dashed line and suggested by \cite{lazarian_2025}. 

The square root scaling found in the incompressible simulations of weak turbulence by \cite{lazarian_2025} is reminiscent of strongly magnetized and compressible turbulence \citep{federrath_2016.magnetic.field.amplification,beattie_2020,beattie_2022_energy_balance,skalidis_tassis2021,skalidis_2021_bpos,skalidis_2023.analytical}. The energetics of highly compressible turbulence are dominated by nonpropagating, large-scale coherent structures \citep{yang_2019.compressible.turb.structures.energetics}. Using first-principles calculations, \cite{skalidis_2023.analytical} showed that the dynamics of these structures naturally lead to the scaling $\delta u \sim \sqrt{B_0 \delta B}$, as a consequence of the strong magnetic compressibility measured by the variance of $\vec{\delta B} \cdot \vec{B}_0$. 

Obtaining a square root scaling in incompressible turbulence simulations is likely an effect of the incompressibility constraint -- $\vec{\nabla} \cdot \vec{u} =0$ and $\vec{\nabla} \cdot \vec{\delta B} = 0$ -- or the forcing function. Velocity (kinetic) forcing in incompressible simulations artificially correlates oppositely traveling wave packets \citep{maron_2001}. Additionally, driving turbulence faster than the eddy relaxation time can lead to changes in the spectral scalings \citep{mason_boldyrev_2008.various.forcing.turb.spectrum}. Simulations of strong incompressible turbulence lead to balanced energy cascades, thereby complying with $\delta u \sim \delta B$ \citep{bian_2019.scale.dependent.prandtl}. 

We conclude that the square root scaling obtained in the incompressible MHD simulations of \cite{lazarian_2025} is either a property of weak turbulence only or an artifact related to the driving mechanism. On the other hand, the scaling $\delta u \sim \sqrt{B_0 \delta B}$ has been verified by several highly compressible numerical simulations with various driving mechanisms and codes \citep{skalidis_2021_bpos}.

\begin{figure*}
   \centering
   \includegraphics[width=\hsize]{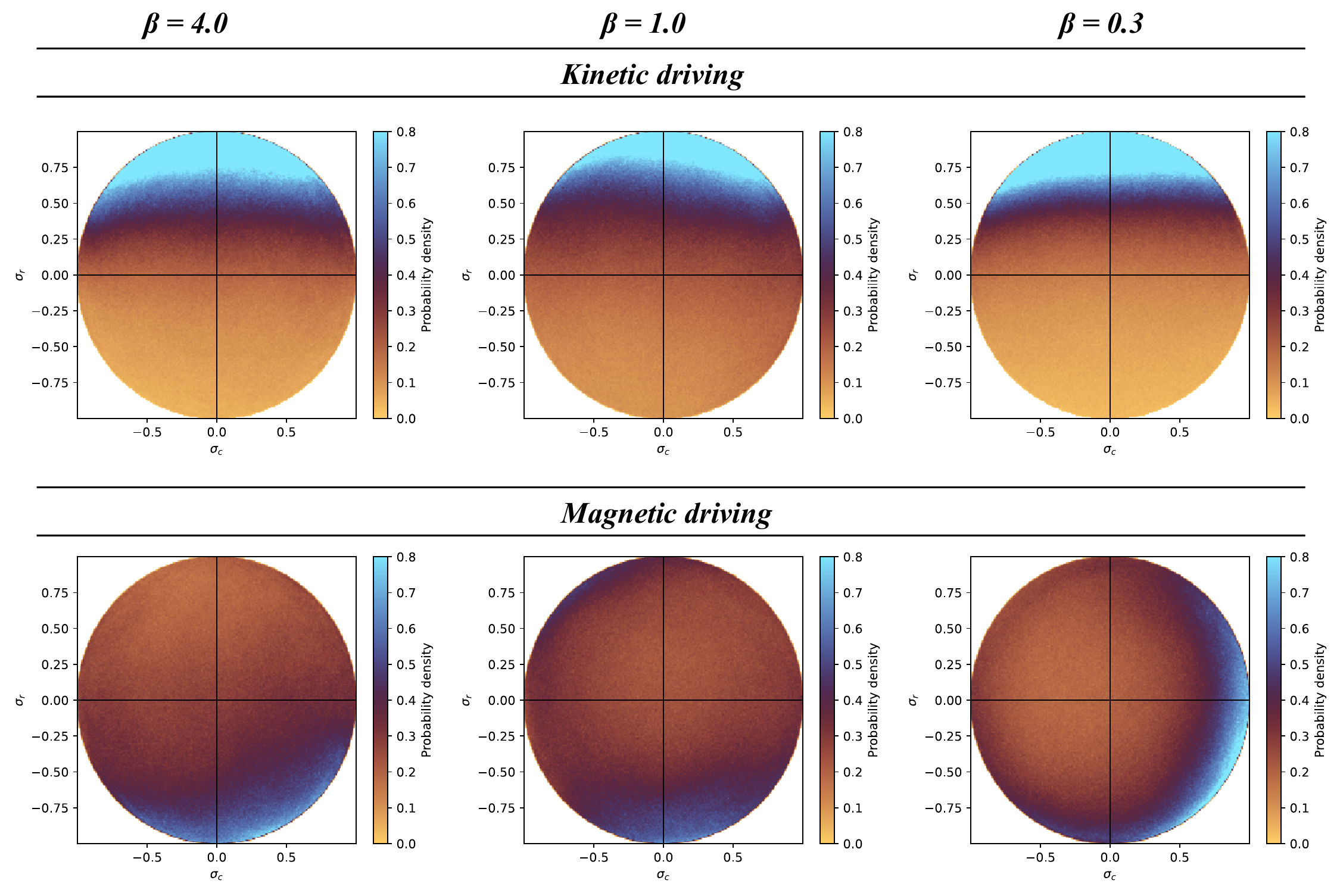}
    \caption{Joint probability density functions of residual energy and cross-helicity for the two driving schemes. Magnetically driven turbulence excited dynamically aligned fluctuations, consistent with the solar wind turbulence observations. Kinetically driven turbulence leads to weak alignment between velocity and magnetic fields, underscoring the non-Alfv\'enic nature of this turbulence, induced by density inhomogeneities.}
    \label{fig:cross_hel_residual_plane}
\end{figure*}
   
\subsection{Scale-dependent energetics}
\label{sec:scaling_properties}

We investigated the influence of injected modes on the resulting energy cascade. Fig.~\ref{fig:power_spectra} shows the kinetic (purple solid curves) and magnetic (yellow solid curves) power spectra (Eq.~\ref{eq:power_spectrum}) of the simulations compensated by different scalings, as indicated in the labels of the vertical axes. The top row shows turbulence simulations driven kinetically, while the bottom row shows simulations where turbulence is driven magnetically. From left to right, we can see that the plasma beta decreases. We note that \Ms\ $\sim 0.1$ in all simulations and, hence, the $\beta$ variations correspond to differences in $|\vec{B_0}|$ or \MA.

Using the Helmholtz decomposition method, we decomposed the kinetic energy spectrum into a solenoidal and a compressible component to explore how their relative ratios change with $\beta$ and forcing. According to Helmholtz decomposition, any smoothly varying field can be decomposed into divergence-free (incompressible) and curl-free (compressible) modes, expressed as
\begin{align}
	& \vec{\hat{u}} = \vec{\hat{u}}_c + \vec{\hat{u}}_s,  \\
	& |\vec{\nabla} \times \vec{\hat{u}}_c| = 0, \\ 
	& \vec{\nabla} \cdot \vec{\hat{u}}_s = 0,
\end{align}
where $\vec{\hat{u}}_s$ and $\vec{\hat{u}}_c$ correspond to the solenoidal and compressible components of the velocity field. A detailed explanation of this decomposition method can be found in the literature \citep[e.g.,][]{hu_2002.helmholtz_decomposition, semadeni_2024.turb.driven.gravity, 
connor_2025.SN.turbulence}.

Incompressible spectra are shown as purple dashed curves in Fig.~\ref{fig:power_spectra}, while compressible spectra are shown as dashed-dotted curves. We discuss the spectral properties of turbulence separately for each driving method below. 

\subsubsection{Magnetic driving: Zero residual energy}
\label{sec:scaling_magnetic_driving}

Both the kinetic and magnetic energy spectra of magnetically driven turbulence show a prominent Iroshnikov - Kraichnan cascade, $E(k) \propto k^{-3/2}$  \citep[IK,][]{Iroshnikov_1964, kraichnan_1965.mhd.turb.inertial.range}. We do not refer to the IK cascade with the strict definition of the term because the initial phenomenology assumes isotropic fluctuations, while strongly magnetized turbulence (as in our simulations, where \MA~< 1) is highly anisotropic \citep{shebalin_1983.mhd.anisotropy,higdon_1984,cho_vishniac_2000.turbulence.anisotropy,oughton_2013}. Instead, we employed the more general use of the term for any spectrum following $k^{-3/2}$ \citep[e.g.,][]{boldyrev_2005.dynamic.alignment}. Our numerical results are consistent with past numerical simulations regarding the scaling properties of the inertial range of Alf\'enic turbulence \citep{muller_grappin_2025.power_spectra.incompressible.MHD,perez_boldyrev_2009.cross.helicity.role}, as well as the subsonic cascade in recent extreme resolution MHD turbulence simulations \citep{Beattie2025_ISM_cascade_nature_astro}. 

The magnetic spectra are dominated by incompressible fluctuations. The relative contribution of compressible modes to the total energy depends on $\beta$. In Table~\ref{table:sim_summary}, we show the relative energy ratio between compressible and incompressible modes ($E_c/E_s \sim u^2_c /u^2_s$). $E_c/E_s$ increases as $\beta$ decreases from 4.0 to 1.0, but the further reduction of $\beta$ to 0.3 is not accompanied by a corresponding increase in the energy carried by the compressible modes. Magnetically driven simulations show a well-developed compressible-mode kinetic spectrum for all $\beta$ (Fig.~\ref{fig:power_spectra}).

The compressible modes power spectrum (dashed dotted line in Fig.~\ref{fig:power_spectra}) is characterized by the same scaling (-3/2) as the incompressible modes. Given the low values of $E_c/E_s$, we conclude that  compressible modes passively follow the cascade of Alfv\'en waves \citep{schekochihin_2009.reduced.mhd.analytical}. This is evident by the slow mode evolution (Eq. 22 in \citealt{schekochihin_2009.reduced.mhd.analytical}, also \citealt{Bhattacharjee_1988,Bhattacharjee_1998}) which to leading order yields that gas pressure fluctuations evolve as: $\delta p \sim - \delta B_\parallel B_0/(4\pi)$. For an isothermal equation of state, $\delta p = c_s^2 \delta \rho$, we obtain
\begin{equation}
   \label{eq:slow_mode}
   \langle \delta \rho^2 \rangle^{1/2} \sim \langle \delta B_\parallel^2 \rangle^{1/2} B_0 / \left(4\pi c_s^2 \right).
\end{equation}  

In Fig.~\ref{fig:slow_mode}, we illustrate the pressure balance relation (Eq.~\ref{eq:slow_mode}) as a function of plasma beta. Black points correspond to magnetically driven simulations. Overall, the pressure balance relation is satisfied with deviations smaller than a factor of two for every case, but $\beta \geq 1$ simulations show greater consistency with the theoretical relation. The transition happening at $\beta=1$ marks the dominance of magnetic over thermal energy in low-plasma beta fluids.  

From Fig.\ref{fig:power_spectra}, it is evident that no net residual energy develops in the inertial range of magnetically driven simulations. The energy cascade is equally distributed between kinetic and magnetic energies at all scales except the forcing scale ($k_f \sim 1.5$). We conclude that magnetically driven turbulence leads to a balanced cascade with \Eresidual$\approx 0$ throughout the inertial range. This is the second result of this analysis. 

Similarly balanced spectra have also been reported in selected solar wind intervals \citep[e.g.,][]{Kasper_2021.PSD.near.SUn.phrl}. However, the residual energy properties in the solar wind are not uniform, and most frequently reveal a significant magnetic excess \citep{chen_2013.residual_energy.solar.wind}. The origin of this discrepancy may be related to physical processes that are absent from the present simulations, such as flow expansion, directional forcing, and other effects related to background inhomogeneities, such as density and temperature gradients. These processes can generate and sustain couplings between the Elsasser fields, which can modify the partition between kinetic and magnetic energies \citep{yang_2022.energy.transfer.elsasser.coupling}. The role of the Elsasser-field couplings and their connection to the shallow spectra obtained in the kinetically driven simulations are further discussed in Section~\ref{sec:theoretical_consideration_kinetic}.

\subsubsection{Kinetic driving: positive residual energy}
\label{sec:scaling_kinetic_driving}

The top row in Fig.~\ref{fig:power_spectra} shows the power spectrum properties of kinetically driven turbulence. The scaling properties of kinetically driven turbulence are close to -1 for both magnetic and kinetic power spectra, which is significantly different from magnetically driven simulations. Inertial ranges shallower than $k^{-3/2}$ have been reported in the literature by several authors \citep{cho_2002.viscous.damped.turbulence,xu_2022.local.turb,lazarian_2025, grete_2021.mhd.spectra.super.alfvenic}. Despite substantial differences among these numerical setups (including magnetization and compressibility level), shallow spectra can appear in a variety of configurations. 

Using the PENCIL code, \cite{haugen_brandenburg_2004.large.scale.dynamo.supression} found a $k^{-1}$ inertial range in weakly compressible turbulence with strong magnetic field, consistent with the results obtained here. A shallower spectrum is expected in the presence of mean field, which has a positive contribution to the magnetic energy at large scales, while at smaller scales its contribution becomes negative. For initially weak fields, \cite{haugen_brandenburg_2004.large.scale.dynamo.supression} found a Kolmogorov cascade. However, they found that even moderate initial mean fields lead to shallow cascades ($k^{-1}$) due to interactions between the mean and fluctuating field, which is also supported by \cite{kleeorin_1994.mean.field.shallow.turb}. This result was confirmed in dynamo DNS \citep{brandenburg_1996.magnetic.structures.dynamo}, where strong magnetic tension suppresses kinetic energy transfer \citep{grete_2021.mhd.spectra.super.alfvenic}. 
 
In solar wind turbulence, there is ample evidence of $k^{-1}$ cascades. Voyager has measured time-varying spectra whose power scales with frequency as $1/f$. According to the Taylor hypothesis\footnote{The frequency $\omega$ of a wave mode with wavevector $\vec{k}$ can be transformed from the plasma to spacecraft (sc) rest frame as $\omega_{sc} = \omega + \vec{k} \cdot \vec{v_{sc}}$; $\vec{v_{sc}}$ is the advection velocity of the wave with respect to sc. For super-Alfv\'enic motions, which are typical in the interplanetary space, $\omega_{sc} \approx \vec{k} \cdot \vec{v_{sc}}$. Violations to this hypothesis may arise under certain conditions  \citep{howes_2014.taylor.hypothesis.break}.}, wave mode temporal frequencies $f$ can be translated to spatial frequencies $k$. In this case, the $1/f$ temporal signal corresponds to a $k^{-1}$ spatial scaling. 

The $1/f$ range (or flickering noise) stores the majority of the turbulent energy at large scales in solar wind turbulence \citep{matthaeus_1986.1f.noise}. This energy-containing range of the turbulence cascade seems to transition to a direct cascade with typical scalings, such as -5/3 or -3/2 \citep{mondal_2025.1/f.two.subinertial.ranges}. Interactions of Alfv\'en waves scattered off from background density inhomogeneities (reflection-driven turbulence) explain the origin of the $1/f$ range \citep{velli_1989.reflection.driven.turb,magyar_2022.cascade.noise.properties,meyrand_2025.reflection.driven.turb}. 

Incompressible fluctuations dominate the kinetic spectra (black dashed lines) of the kinetically driven simulations. Unlike magnetically driven turbulence, for $\beta = 4.0$ the inertial range of the compressible-modes power spectrum has not developed. Interestingly, the spectrum of compressible ($\propto k^{-3/2}$) is significantly different from incompressible ($\propto k^{-1}$), suggesting that their corresponding interactions are governed by different processes. As with magnetically driven simulations, we see that kinetically driven simulations satisfy the slow mode pressure balance relation (Fig.~\ref{fig:slow_mode}), indicating that density perturbations arise from slow modes. The relative energy contribution of compressible modes is lower in kinetically  than magnetically driven simulations. 

The power spectra analysis strongly suggests an imbalanced energy exchange, which translates into a net and positive residual energy cascade (Fig~\ref{fig:power_spectra}). A similar result has been found by \cite{haugen_brandenburg_2004.large.scale.dynamo.supression} who attributed this excess to the work exerted by the mean field.

Fig.~\ref{fig:er_cascade} visualizes the power spectrum of \Eresidual,  which is always positive in these simulations, compensated by different power laws, $k^{\alpha}$, shown in the legend. The left, middle, and right panels correspond to $\beta = 4.0, 1.0, $ and 0.3, respectively. The inertial range of \Eresidual\ extends from approximately $k \sim 3$ to $k \sim 50$. The scaling ($\alpha$) of \Eresidual\ decreases with $\beta$. For $\beta > 1$, $\alpha$ is within -5/3 and -2, for $\beta \approx 1$, $\alpha$ is within -5/3 and -3/2, while for $\beta < 1$, $\alpha \approx -1$. This is the third major outcome of this work.

Differences in the energy mode content \citep[e.g.,][]{zank_matthaeus_1993.nearly.incompressible.turbulence.perturbative.models} explain the dependence of $\alpha$ on $\beta$. The obtained range, $\alpha ~ \epsilon ~ [-1, -2]$ is consistent with literature values \citep[e.g.][]{boldyrev_2011.spectral_scalings.solarwind,wang_2011.residual.energy.weak.turb.spotaneosuly.generated,gogoberidze_2012.residual.energy.incompressible.EDQNM}, suggesting that the scaling variance among these works might come from $\beta$. However, we note that in our simulations \Eresidual\ > 0, while in most of the aforementioned works, \Eresidual\ < 0. A $\beta$ dependence on the scaling of \Eresidual\ was also noted in the highly compressible turbulence simulations of \cite{Vogel2011.thesis}. 

\section{Theoretical considerations}
\label{sec:theory}

Our simulations are weakly compressible with $10\% \lesssim E_c/E_s \lesssim  25\%$; except for the kinetically driven simulations with $\beta = 4.0$ for which $E_c/E_s \sim 2 \%$. Even in the presence of inhomogeneities, turbulence can show a dominant  Alfv\'enic behavior \citep{magyar_2019.elssaser.compressible.MHD}, although deviations are expected. Here, we delve into the physical picture of the turbulence induced by each driving mechanism.
 
 \subsection{Magnetically driven simulations}
\label{sec:theoretical_consideration_magnetic}

Our magnetically driven turbulence simulations are characterized by a $\propto k^{-3/2}$ cascade, which is consistent with Alfv\'enic turbulence. The main features of this turbulence are summarized below, as well as a comparison with the numerical data. 

\subsubsection{Theory of dynamically aligned turbulence}

We express the incompressible MHD equations, which can be expressed in terms of the Elsasser variables,
\begin{equation}
	\label{eq:incomp_MHD}
	\frac{\partial \vec{z}^{\pm}}{\partial t} \pm \left( \vec{V_A} \cdot \vec{\nabla} \right  ) \vec{z^{\pm}} + \left( \vec{z^{\mp}} \cdot \vec{\nabla} \right) \vec{z^{\pm}} = - \vec{\nabla} P,
\end{equation}
where $z^{\pm} = u \pm \delta B/\sqrt{4\pi \rho}$, $V_A = B /\sqrt{4\pi \rho}$, and $P$ is the gas pressure, which ensures that the incompressibility condition is satisfied. Density ($\rho$) is by definition constant in incompressible turbulence. 

When any of \zplus\ or \zminus\ are zero, the nonlinear term vanishes, $\left( \vec{z^{\mp}} \cdot \vec{\nabla} \right) \vec{z^{\pm}} =0$, reducing Eq.~(\ref{eq:incomp_MHD}) to a wave equation for the other variable. For example, if we set $z^-=0$, then Eq.~(\ref{eq:incomp_MHD}) becomes $\partial_t \vec{z^+} +  \left( \vec{V_A} \cdot \vec{\nabla} \right  ) \vec{z^{+}} = 0$, whose solution is a wave propagating along the mean field. The same applies for $z^+=0$, but in that case the waves propagate in the opposite direction. This leads to the widely accepted conclusion that incompressible turbulence arises due to interactions between oppositely traveling Alfv\'en waves.

The IK turbulence cascade scales as $k^{-3/2}$, but the isotropy assumption employed in this phenomenology does not apply when strong (mean) magnetic fields are present \citep{shebalin_1983.mhd.anisotropy,higdon_1984}. The anisotropy of turbulence was considered by Goldreich \& Sridhar \citep[GS,][]{sridhar_1994,goldreich_1995}, although they predicted a $k^{-5/3}$ cascade. 

According to GS, the energy cascade rate $\epsilon$, which is constant, scales as $u_\ell^2/\tau \sim \epsilon$, where $\tau$ is the timescale of nonlinear interactions. The GS scaling can be obtained by assuming that $\tau \sim \ell/u_\ell$, which yields that $u_\ell \sim \epsilon~\ell^{1/3}$. The general form of the energy cascade is 
\begin{equation}
	\label{eq:gs_energy_spectrum}
	E(k) \propto \frac{u_k^{2}}{k},
\end{equation}
where $k \sim 1/\ell$. In GS turbulence, we obtain $E(k) \propto k^{-5/3}$.

We can see that DNS with a strong magnetic field yield an anisotropic spectrum that is consistent with GS and a -3/2 scaling, which is, on the other hand, inconsistent with GS \citep{muller_grappin_2025.power_spectra.incompressible.MHD}. \cite{boldyrev_2005.dynamic.alignment} reconciled this tension by postulating that nonlinear interactions, whose timescales are of the form $\tau \sim \ell/u_\ell~\left( V_A/u_\ell \right)^\alpha$, attenuate at smaller scales due to the tendency of magnetic and velocity fluctuations to align. This dynamic alignment is achieved through a process known as turbulence Alfv\'enization, which has been observed in the Sun \citep{grappin_1982.alignment}, and leads to an IK scaling when $\alpha = 1$ \citep{boldyrev_2006.dynamic.alignment}. 

The energy cascade becomes imbalanced when there is a dominant Elsasser variable (e.g., $|z^+|^2 \gg |z^-|^2$) breaking the symmetry in Eqs.~(\ref{eq:incomp_MHD}). The imbalance is quantified by the normalized cross-helicity
\begin{align}
	\label{eq:cross_hel_Elsasser}
	\sigma_c \equiv  \frac{|z^+|^2  - |z^-|^2 }{|z^+|^2  + |z^-|^2},
\end{align}
which can be also written as
\begin{equation}
	\label{eq:cross_hel_comp}
	\sigma_c \equiv 2 \frac{\vec{u} \cdot \vec{\delta B}/\sqrt{4\pi \rho}}{u^2 + \delta B^2/4\pi\rho}.
\end{equation}
The solar wind turbulence is dominated by $z^+$ modes, which are injected by the Sun and propagate outwards; hence, we have $\sigma_c >0$ \citep{goldstein_1995.review.solar.wind.turb}. 

The properties of imbalanced cascades have been a matter of debate \citep[e.g.,][]{lithwick_goldreich_2001.compressible.turb}. In an attempt to mitigate these controversies, \cite{perez_boldyrev_2009.cross.helicity.role} appealed to the scale-dependent (dynamic) alignment theory \citep{boldyrev_2005.dynamic.alignment}. In balanced cascades, \zplus\ and \zminus\ have comparable energy fluxes: $\left( z^+ \right ) ^2/\tau_\ell \sim \left (z^- \right )^2/\tau_\ell \sim \epsilon$, where $\tau_\ell \sim 1/z^{\pm}_\ell k_\perp \theta_\ell$ and $\theta_\ell \propto \ell^{1/4}$ \citep{boldyrev_2006.dynamic.alignment,mason_2006.dynamic.alignment.exponent.mhd.forced,Chandran2015_alignment,Beattie2025_scale_depedenent_aligment}. In imbalanced cascades, the energy fluxes of \zplus\ and \zminus, denoted as $\epsilon^+$ and $\epsilon^-$, respectively, are different. Scale-dependent alignment imposes the following geometrical constraint for local interactions: $z^+ \theta^+_{\ell} \sim z^- \theta^-_{\ell}$, implying that $\tau_{\ell}^+ \sim \tau_{\ell}^-$. Thus, we derive the following cascade ratio for \zplus\ and \zminus: $\left( z^- \right ) ^2/ \left( z^+ \right ) ^2 \sim \epsilon^+ / \epsilon^-$ \citep{perez_boldyrev_2009.cross.helicity.role}. If both $ \epsilon^+$ and $ \epsilon^-$ are constant in time, \zplus\ and \zminus\ have the same scalings, but different normalizations. Thus, an overall balanced cascade ($\langle \sigma_c \rangle = 0$) consists of locally imbalanced patches \citep{Servidio2008_relaxation_in_decaying_turb,perez_boldyrev_2009.cross.helicity.role,Pecora2023_alignment_in_magnetosheath,Beattie2025_scale_depedenent_aligment}. 
 
\subsubsection{Comparison with numerical data}

The aforementioned picture is consistent with our magnetically driven simulations. Fig.~\ref{fig:corss_hel_res_snapshots} shows perpendicular-to-the-mean-field planes of normalized cross-helicity ($\sigma_c$, Eq.~\ref{eq:cross_hel_comp}) and residual energy ($\sigma_r$, Eq.~\ref{eq:residual_energy}); $\vec{B_0}$ points towards the reader. 

Magnetically driven turbulence shows prominent curl-like structures, which is suggestive of Alfv\'en wave propagation. $\sigma_c$ and $\sigma_r$ correlate, suggesting that cross-helicity is a key metric for the balance between magnetic and kinetic energies. For regions with $\sigma_c \approx 1 \Rightarrow |z^+| \gg |z^-|$, while for $\sigma_c \approx -1 \Rightarrow |z^+| \ll |z^-|$. Blue regions show Alfv\'enic fluctuations that propagate towards the reader, while in brown regions fluctuations propagate away from the reader. Local fluctuations in $\sigma_c$  are strong and accompanied by $\sigma_r$, but the energy cascade is overall balanced, as shown by the power spectrum analysis (Fig.~\ref{fig:power_spectra}), consistent with the dynamic alignment theory \citep{boldyrev_2005.dynamic.alignment,boldyrev_2006.dynamic.alignment}.

Figure ~\ref{fig:cross_hel_residual_plane} shows the 2D joint probability density function (PDF) between $\sigma_c$ and $\sigma_r$. Here, the PDFs are circular because both $\sigma_c$ and $\sigma_r$ are normalized such that their maximum values are unity. At the center of the distribution, we have $\sigma_c = \sigma_r = 0$, which corresponds to locally balanced fluctuations. Near the edges, the fluctuations are imbalanced.

In magnetically driven turbulence simulations, the maximum of the PDF is in the fourth quadrant. Because the simulations include explicit viscous terms, cross-helicity is not conserved. Specifically, $\langle \sigma_c \rangle \approx 0$ for the $\beta \gtrsim 1$ runs, while $\langle \sigma_c \rangle \approx 0.13$ for the $\beta=0.3$ run. This difference is also reflected in the PDF of the $\beta< 1$ simulation, whose peak shifts towards the first quadrant. In addition, the probability of positive $\sigma_r$, corresponding to \Ekin~>~\Emag, is enhanced relative to the $\beta \geq 1$ simulations. Overall, the depicted PDFs suggest a mixture of $\sigma_c$ fluctuations and a predominance of magnetic-energy-dominated states ($\sigma_r <0$). This statistical tendency is qualitatively similar to that commonly observed in solar wind turbulence \citep{Bavassano_1998.residual.crosshel.5au,perri_balogh_2010.crosshel.residual.Ulysses,wicks_2013.residual_energy.cross_helicity.energy.containing,popescu_2016.cross.hel.res.energy.ulysses,wu_yang_2024.elssaser.coherence.res.crosshel,chen_2013.residual_energy.solar.wind}. We emphasize the fact that this comparison concerns the distribution of local fluctuations and not the scale-dependent properties.

\begin{figure}
   \centering
   \includegraphics[width=\hsize]{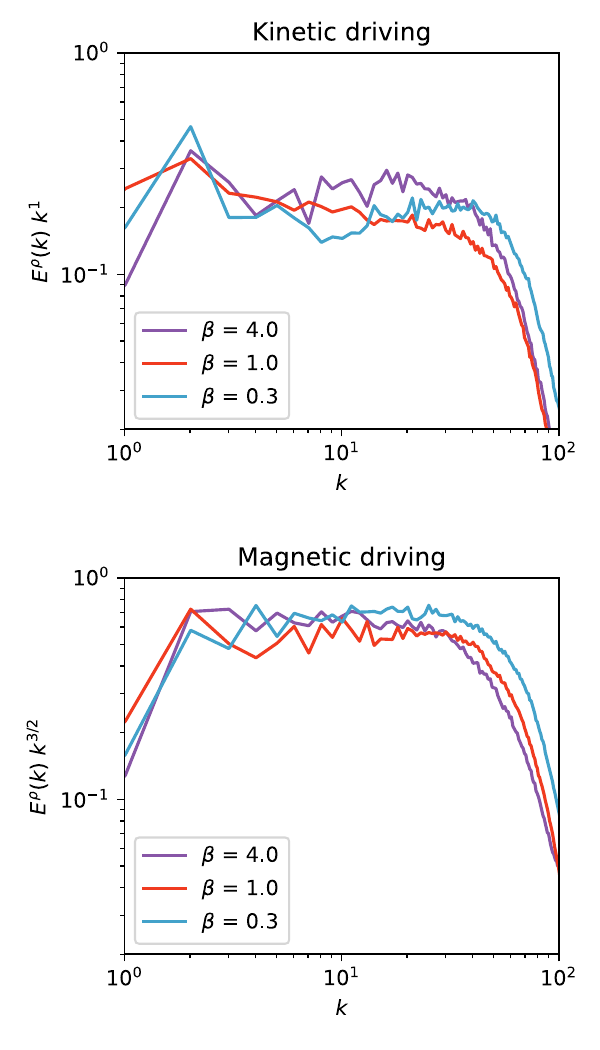}
   \caption{Density power spectra compensated by different scalings as indicated in the labels. Magnetic driving leads to passively mixed density modes, which acquire the IK spectrum of Alfv\'en waves. Kinetic driving leads to a smooth mixing of density perturbations and has the same scaling as passive scalar turbulence. Then density inhomogeneities distort the propagation of the Alfv\'en waves, which acquire such a shallow spectrum.}
   \label{fig:density_PSD}
\end{figure}

\begin{figure}
   \centering
   \includegraphics[width=\hsize]{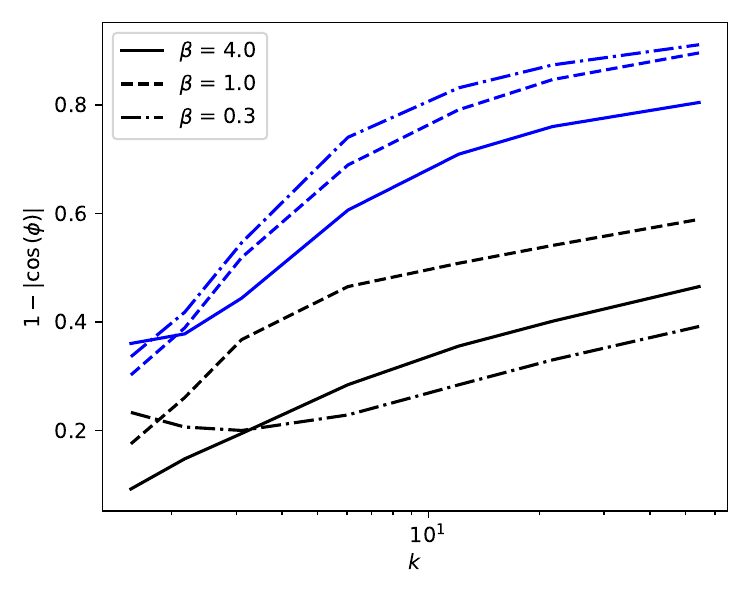}
      \caption{Normalized coherence (Eq.~\ref{eq:projection_angle}) as a function of scale. The various curves correspond to simulations with different plasma beta values. In magnetically driven simulations, shown as blue curves, the coherence metric progressively converges to unity, which implies that the resulting interactions dissipate any Elsasser field correlations injected by the forcing mechanism at the largest scales. In contrast, in kinetically driven simulations, coherence remains significant for the entire range of $k$. This is consistent with the scaling differences between the two driving mechanisms.}
         \label{fig:projection_angle}
\end{figure}

\subsection{Kinetically driven simulations}
\label{sec:theoretical_consideration_kinetic}

Our kinetically driven turbulence simulations have a $k^{-1}$ cascade (Fig.~\ref{fig:power_spectra}). The general cascade form of dynamically aligned turbulence is $E(k) \propto k^{-\left(5 + p \right)/\left(3 + p \right)}$ \citep{boldyrev_2005.dynamic.alignment}. This is fundamentally different from the spectral properties of kinetically driven simulations because there is no $p$ leading to $k^{-1}$. 

According to Eq.~\ref{eq:gs_energy_spectrum}, a shallow $k^{-1}$ scaling implies that mode interactions are weak; hence, their amplitudes scale independently, $u_k \sim \rm const$. Interactions are suppressed when the Elsasser fields are correlated, $\vec{z}^+ \cdot \vec{z}^- \neq 0$ \citep{yang_2022.energy.transfer.elsasser.coupling}. In-situ measurements of solar wind turbulence reveal significant scaling differences between correlated and uncorrelated fluctuations. Large-scale, weakly interacting, correlated modes produce shallower spectra compared to maximally interacting, uncorrelated modes \citep{wicks_2013.prl,wicks_2013.residual_energy.cross_helicity.energy.containing}.

The normalized coherence between the Elsasser fields is a key parameter that determines the cascade energy rate, defined as \cite{wicks_2013.residual_energy.cross_helicity.energy.containing} via
\begin{equation}
	\label{eq:projection_angle}
	\cos \left( \phi \right) = \frac{\vec{z}^+ \cdot \vec{z}^-}{|\vec{z}^+||\vec{z}^-|}.
\end{equation}
Here, when $\phi = 90\degr$, the shearing distortion between two interacting modes is maximized, while when $\phi = 0\degr$ it is minimized.

Figure\ref{fig:projection_angle} shows $1-|\cos(\phi)|$ as a function of $k$ for the magnetically  and kinetically driven simulations. Different line styles correspond to different $\beta$ values. When $1-|\cos(\phi)|=0$, the Elsasser fields are maximally correlated, whereas $1-|\cos(\phi)|=1$ corresponds to orthogonal Elsasser fields.

At the largest scales, $k \sim k_f$, the Elsasser fields exhibit a finite correlation inherited from the driving. While the correlation is somewhat stronger in the kinetically driven simulations, the principal difference lies in its subsequent evolution. In the magnetically driven simulations, the Elsasser fields were shown to progressively decorrelate as the cascade develops, approaching an almost orthogonal state at small scales. In contrast, the kinetically driven simulations retained a significant fraction of the injected correlation across all scales, indicating the presence of long-lived couplings between the Elsasser fields.

This difference is also reflected in the volume-averaged energy partition of the turbulence. In the magnetically driven simulations, where the Elsasser correlation is weaker, the volume-averaged kinetic and magnetic energies remain closer to equipartition (\alfvenratio~$\sim 1$) than in the kinetically driven simulations, which retain stronger correlations and exhibit systematically larger Alfv\'en ratios (\alfvenratio~$\sim 3$; Fig.~\ref{fig:ra_beta}; Table~\ref{table:sim_summary}).

The  presence of stronger Elsasser correlations suggests that a significant fraction of the injected fluctuations interact only weakly and therefore preserve their large-scale correlations throughout the cascade, or that a mechanism continuously regenerates coherence between the Elsasser fields. The former interpretation is consistent with the presence of velocity–dominated coherent (weakly–interacting) structures. The latter possibility naturally motivates the consideration of inhomogeneities in turbulence dynamics \citep[e.g.,][]{velli_1993.reflection.driven,yang_2022.energy.transfer.elsasser.coupling}. These scenarios are not mutually exclusive, since large-scale coherent structures can provide the slowly varying (inhomogeneous) environment where Alfvénic interactions evolve.

\subsubsection{Theory of reflection-driven turbulence}

Large-scale inhomogeneities provide a natural mechanism for generating correlations between the Elsasser fields. One such example is reflection-driven turbulence, where Alfv\'en waves are scattered by inhomogeneities, leading to coupling between counter-propagating fluctuations \citep{velli_1989.reflection.driven.turb,velli_1993.reflection.driven}. When the Alfv\'en speed is nonuniform, the propagation of the Elsasser variables is governed by
\begin{align}
	\label{eq:reflection_elssaser}
	\frac{\partial \vec{z}^{\pm}}{\partial t} \pm \overbrace{ \left( \vec{V_A} \cdot \vec{\nabla} \right  ) \vec{z^{\pm}} + \left( \vec{z^{\mp}} \cdot \vec{\nabla} \right) \vec{z^{\pm}}}^{\rm homogeneous}  \mp \qquad\qquad  \\
	\qquad\qquad\underbrace{ \left(\vec{z}^{\mp} \cdot \vec{\nabla} \right) \vec{V_A} +  \frac{1}{2} 
 \left(\vec{z^{-}} - \vec{z^{+}} \right)  \left(\vec{\nabla} \cdot \vec{V_A} \right)}_{\rm inhomogeneous} 
		= - \vec{\nabla} P.
\end{align}
Here, the inhomogeneous terms proportional to $\nabla V_A$ linearly couple the two Elsasser fields and enable the generation of a minority component even in the absence of direct nonlinear interactions.

Inhomogeneities excite counter-propagating waves, usually referred to as ``anomalous.'' In this case, the Elsasser variables consist of a classical component directly generated by the forcing, along with an anomalous component, arising from reflections. In reflection-driven turbulence, nonlinear interactions are dominated by these anomalous fluctuations, which remain phase-correlated with their parent wave and therefore shear it coherently. This is in contrast to the classical Alfv\'enic turbulence, where nonlinear interactions arise from independently propagating counter-propagating waves and are consequently less coherent.
  
The energy flux rate of a parent mode, for example, \zplus\ in the solar wind, is $\epsilon^+ \sim \left( z^+ \right)^2/\tau^+$. Interactions are due to anomalous \zminus; hence, we have $\tau^+ \sim 1/\left( k z^- \right)$ \citep{velli_1989.reflection.driven.turb}. Intuitively, we would expect that the reflected wave carries a fraction ($q$) of the parent wave energy, which is a function of the density contrast \citep[e.g.,][]{stein_1971.reflection.mhd}, over some distance with respect to the coherence length of Alfv\'en waves ($\ell_A$). Thus, $z^- \sim q z^+ (\ell/\ell_A)$, where $\ell_A \sim V_A T_A$ and $T_A$ is the Alfv\'en time \citep{velli_1989.reflection.driven.turb}. These relations lead to an energy rate for \zplus\ independent of scale: $\epsilon^+ \sim q \left(z^+\right)^3/(V_A T_A)$, leading to $E(k) \propto k^{-1}$. The same scaling also applies for anisotropic fluctuations \citep{perez_2013.reflection.driven.anisotropy}. 
 
In the imbalanced limit, for example, when $|z^+| \gg |z^-|$, the linear propagation of the Elsasser variables to leading order becomes
\begin{align}
\frac{\partial \vec{z}^+}{\partial t} +  \left(\vec{V}_A\cdot\nabla\right)\vec{z}^+
&\approx
\frac{1}{2}\vec{z}^+
\left(\nabla\cdot\vec{V}_A\right)
-\nabla P, \\
\frac{\partial \vec{z}^-}{\partial t}
-
\left(\vec{V}_A\cdot\nabla\right)\vec{z}^-
&\approx
-\left(\vec{z}^+\cdot\nabla\right)\vec{V}_A
+
\frac{1}{2}\vec{z}^+
\left(\nabla\cdot\vec{V}_A\right)
-\nabla P .
\end{align}

The propagation of \zplus\ will experience an amplitude modulation due to changes in the Alfv\'en speed. On the other hand, the propagation of \zminus\ is linearly coupled to \zplus, due to the reflection terms. The amplitude of \zminus\ will grow but not indefinitely. When \zplus~$\sim$~\zminus, nonlinear interactions become important, while $\left(\vec{z^{-}} - \vec{z^{+}} \right)  \left(\vec{\nabla} \cdot {V_A} \right) \rightarrow 0$. The reflection term, $\left(\vec{z}^{\mp} \cdot \vec{\nabla} \right) \vec{V_A}$, couples the two Elsasser variables, making the system of \zplus\ and \zminus\ equations symmetric. Thus, contrary to the standard dynamically aligned Alfv\'enic turbulence that is locally imbalanced. This is consistent with our magnetically driven simulations. Furthermore, density inhomogeneities tend to equalize the energy contained in \zplus\ and \zminus. As a result, for any closed system, we expect that $\sigma_c \approx 0$, even if initial fluctuations are imbalanced. In solar wind turbulence, \zplus\ $\gg$ \zminus\ because the Sun predominantly injects \zplus\ fluctuations, thereby maintaining a significant level of imbalance. 

We also note that local inhomogeneities can also generate coupled Elsasser fields with a similar impact on the cascade: decreasing the spectral scaling to -1 \citep{yang_2022.energy.transfer.elsasser.coupling}. All of these mechanisms involve some linear \citep[e.g.,][]{velli_1993.reflection.driven} or nonlinear \citep[e.g.,][]{yang_2022.energy.transfer.elsasser.coupling} coupling term that is proportional to the spatial gradients of the Alfv\'en speed, but the corresponding rates might differ. It is beyond the scope of the current manuscript to distinguish between the underlying processes; however, in Sect.~\ref{sec:why_forcing_matters}, we provide evidence that the dynamics of the kinetically driven simulations are more significantly influenced by large-scale inhomogeneities than those of the magnetically driven simulations, which might account for the spectral differences.
 
\subsubsection{Comparison with numerical data}

The present analysis does not allow us to conclusively identify a reflection-driven mechanism. However, the statistical and morphological properties of the kinetically  and magnetically driven turbulence differed substantially, indicating that the interactions governing the two regimes are fundamentally different. The upper two rows in Fig.~\ref{fig:corss_hel_res_snapshots} show perpendicular planes of $\sigma_c$ and $\sigma_r$ of our kinetically driven simulations. In the kinetically driven simulations, the normalized residual energy is predominantly positive ($\sigma_r$ > 0), indicating a systematic excess of kinetic over magnetic energy. The corresponding cross-helicity fluctuations are comparatively weak and spatially fragmented, producing a complex mixture of locally correlated and anti-correlated regions. As a result, the turbulence appears more stochastic and less organized than in the magnetically driven case.

The joint probability density functions shown in the top row of Fig.~\ref{fig:cross_hel_residual_plane} further highlight differences with the magnetically driven runs (bottom row). Kinetically driven turbulence occupies a relatively narrow region of the ($\sigma_c$, $\sigma_r$) plane, concentrated around positive residual energy and small cross-helicity values. In contrast, magnetically driven turbulence occupies a much larger fraction of the available parameter space and exhibits significantly stronger cross-helicity fluctuations. These results demonstrate that the driving mechanism strongly influences the statistical organization of the turbulent fluctuations, even when the global \Ms\ and \MA\ of the turbulence remain similar.

\section{Discussion}
\label{sec:discussion}

\subsection{Why forcing matters}
\label{sec:why_forcing_matters}

Here, we consider why magnetic driving results in an Alfv\'enic, while kinetic driving in a shallower cascade. The answer depends on how energy is distributed between the two forcing schemes.

We applied solenoidal, isotropic forcing in all simulations. For magnetic driving, the imposed fluctuating field $\vec{\delta B}$ satisfies $\vec{\nabla} \cdot \vec{\delta{B}} = 0$, so each driven Fourier mode obeys $\vec{k} \cdot \vec{\delta B_k} =0 \Rightarrow k_\perp \delta B_\perp \sim k_\parallel \delta B_\parallel$. If the forcing is (nominally) isotropic in $\vec{k}$-space, populating modes at all angles while also enforcing the solenoidal condition, it tends to project power into perpendicular modes. In this case, forcing acts primarily on $\delta B_\perp$, while parallel motions and density perturbations are not directly forced: they emerge through mode coupling. Thus, density behaves primarily as a passive scalar and the resulting structures have little feedback on the flow.

Velocity driving, by contrast, injects velocity perturbations with both perpendicular and parallel polarizations, and so contains the polarization of both Alfvén and pseudo-Alfvén modes. The incompressible condition $\left( \vec{\nabla} \cdot \vec{u}=0 \right)$ implies $k_\perp u_\perp \sim k_\parallel u_\parallel$ for the driven modes; in an incompressible flow, this relation must be satisfied at all times and can therefore produce case-sensitive results \citep[e.g.,][]{lazarian_2025}. Incompressible velocity forcing drives $u_\parallel$, which is associated with density perturbations. In this case, density structures are not solely a by-product of mode coupling because they are continuously generated by the forcing. The injected large-scale density fluctuations could rather represent weakly interacting structures, providing the spatially inhomogeneous environment through which Alfv\'en waves propagate and interact.

The influence of driving in the density power spectra is evident in Fig.~\ref{fig:density_PSD}. The upper and lower panels show results for the kinetically  and magnetically driven runs, respectively. The density power spectrum in kinetically driven turbulence scales as $k^{-1}$. A similar scaling has been reported in the fast solar wind \citep{bruno_2014.density.psd}. The pronounced peak near the injection scale ($k_f \sim 1.5$) indicates that large-scale density fluctuations are directly injected by the forcing and subsequently advected by the flow. Alfv\'en waves can acquire the scaling of the background density field \citep{magyar_2022.cascade.noise.properties}, which explains the shallow spectral properties of the kinetically driven turbulence simulations (Fig.~\ref{fig:power_spectra}). On the other hand, the density power spectrum in magnetically driven simulations (lower panel) is flat and scales as $k^{-3/2}$, implying that density fluctuations emerge self-consistently by the injected rotational modes due to their coupling with compressible ones \citep{cho_lazarian_2002.low_beta.turb,schekochihin_2009.reduced.mhd.analytical}. 

These differences between the driving schemes are also reflected in the relative ratios between the correlation lengths of the various fields. For kinetically driven turbulence, we find that $\ell_u > \ell_\rho \gtrsim \ell_b$, while for magnetically driven turbulence, this is $\ell_b > \ell_\rho \gtrsim \ell_u$. The evolution timescale of each field ($\tau$) is given by the ratio between the correlation lengths and the characteristic propagation speed, $V_A$.

The field with the longest correlation time acts as a slowly varying background experienced by the emerging fluctuations. In the magnetically driven simulations, $\tau_b > \tau_\rho$, implying that density fluctuations adjust to a magnetic background that is effectively frozen over their lifetime. In contrast, in the kinetically driven simulations $\tau_\rho > \tau_b$, suggesting that magnetic fluctuations propagate through a slowly varying inhomogeneous background, providing conditions favorable for Elsasser-field coupling generation.

\subsection{Driving mechanisms and their impact on turbulence}

Maintaining astrophysical turbulence requires constant energy injection to overcome losses \citep{stone_1998.turb.dissipation,mac_low_1998}. Nature achieves this through large- and small-scale processes. For example, coronal mass ejections or recurring jets drive solar wind turbulence \citep{webb_2012.dmes.obs,soljento_2023.imbalanced.turb.large.scale.shear}, feedback processes drive interstellar medium turbulence \citep{beattie_2025.sne.driven.turb,connor_2025.SN.turbulence,Beattie2026_small_to_large_scales}, and jets from active galactic nuclei can drive turbulence in galaxy halos \citep{fabian_2012.agn.feedback}.

Turbulence-in-a-box simulations, as those presented in this work, employ forcing functions to mimic these energy injection mechanisms. There are two main parameters determining the forcing properties in DNS: its normalization and auto-correlation time ($t_{f}$). The driving function is nominally normalized by the energy dissipation rate \citep[e.g.,][]{stone_1998.turb.dissipation} or a constant amplitude \citep[e.g.,][]{brandeburg_2001.pencil} with turbulence properties being sensitive to this choice \citep{grete_2018.mhd.forcing.superalfvenic}. 

The relative ratio between the auto-correlation time and eddy relaxation time plays a significant role on the turbulence properties \citep{mason_boldyrev_2008.various.forcing.turb.spectrum,yoon_2016.driving.schemes.autocorr,grete_2018.mhd.forcing.superalfvenic,Beattie2026_strouhal_number_of_SN_driven}. If the auto-correlation time of forcing is longer than the eddy relaxation time, the system evolves in a quasi--balanced way; meanwhile for $\delta$-in-time correlations, the fluid is instantaneously perturbed. 

In incompressible DNS, where \Ms~$\rightarrow 0$, short auto-correlation times can steepen the energy cascade from $k^{-3/2}$ to $k^{-5/3}$ \citep{mason_boldyrev_2008.various.forcing.turb.spectrum}. However, in weakly compressible turbulence, the auto-correlation time has a minor effect on the scaling, but it can affect the relative energy ratio between solenoidal and compressible modes \citep{grete_2018.mhd.forcing.superalfvenic}. Driving effects are weaker for \Ms~$\gg 1$ \citep{yoon_2016.driving.schemes.autocorr, grete_2018.mhd.forcing.superalfvenic,skalidis_2021_bpos}.
 
In our simulations, the forcing function is $\delta$-in-time correlated and has a constant normalization \citep{brandeburg_2001.pencil, brandenburg_dobler_2001}. We expect no impact on the obtained scalings by the driving function (Sects.~\ref{sec:scaling_magnetic_driving} and \ref{sec:scaling_kinetic_driving}). However, the choice of $t_f$ could affect the obtained $E_c/E_s$ ratios (Table~\ref{table:sim_summary}) and the cross-helicity distributions (Fig.~\ref{fig:cross_hel_residual_plane}) because the correlation between magnetic field and density fluctuations depends on it \citep{yoon_2016.driving.schemes.autocorr, grete_2018.mhd.forcing.superalfvenic}. 

Our work investigates differences between kinetic and magnetic driving, thereby broadening the parameter space of driving studies \citep[see also][]{lazarian_2025}. Contrary to the auto-correlation time, magnetic and kinetic driving give rise to different cascades even for weakly compressible turbulence. An interesting avenue for future exploration is the influence of auto-correlation time on these two driving schemes.  

\subsection{Positive residual energy and where to find it in interplanetary space}

Modeling the solar wind turbulence is beyond the scope of this manuscript, but our numerical results reproduce several key properties. It is therefore tempting to discuss the implications of our simulations for solar wind turbulence. We emphasize the characteristics of kinetically dominated turbulence and identify where it can occur.

In situ measurements suggest that solar wind turbulence is, on average, dominated by Alfv\'enic interactions. Some key characteristics include: 1) \alfvenratio\ $\approx 0.5$ \citep{tu_marsch_1995.book.waves.structures.solar.wind}; 2) power spectra with scalings approximately equal to -3/2 \citep{podesta_2007.spectral.exponents}; and 3) locally imbalanced fluctuations \citep{sioulas_2025.alfven.damping}. 

Solar wind turbulence properties vary significantly across scales and environments in interplanetary space. \alfvenratio\ decreases from greater then unity to less than unity with the heliocentric distance and frequency \citep[e.g.,][]{tu_marsch_1995.book.waves.structures.solar.wind,podesta_2007.spectral.exponents}. Additionally, the power spectrum scaling is not constant: it transitions from -1 to -3/2 with increasing frequency \citep{mondal_2025.1/f.two.subinertial.ranges}. 

A -1 scaling arises from density inhomogeneities (reflection-driven turbulence). However, even in this range, super-Alfv\'enic simulations predict \Eresidual\ < 0 \citep[e.g.,][]{meyrand_2025.reflection.driven.turb}; \Eresidual\ > 0 is expected in shock-compressed regions \citep{good_2025.pos.residual.energy.shocks},

Figure 1 in \cite{Kasper_2021.PSD.near.SUn.phrl} shows the dominant velocity fluctuations in the low-frequency energy-containing part of the spectrum ($1/f$) in a sub-Alfv\'enic region. Additionally, \cite{bowen_2020.parker.probe.sheat.residual} observed localized patches with \Eresidual\ > 0 in a sheath in the inner heliosphere within 0.5 a.u. of the Sun. These observational constraints are consistent with the properties of our kinetically driven simulations. Based on our results, we expect weak alignment between velocity and magnetic fluctuations in these regions. 

Slow wind streams generally have weaker alignment than fast winds. \cite{shi_2021.Alfvenic.nonAlfvenic.slow.fast.winds} found significant scatter in wind properties, even among streams with comparable propagation speeds. Differences in the injected modes that generated these events could explain the variety in wind-stream properties. For the slow-wind dynamics, our simulations indicate positive residual energy with a $\beta$-dependent scaling (Fig.~\ref{fig:er_cascade}).

\section{Conclusions}
\label{sec:conclusions}

We performed DNS of weakly compressible, mean-field guided MHD turbulence with the PENCIL code to study the properties of the residual energy. We employed both magnetic and kinetic forcing. The sonic Mach number in all simulations is close to 0.1, whereas we considered the following initial magnetic field strength values for each driving method: $B_0 = 0.5, 1.0, 2.0$, corresponding to $\beta$ = 4.0, 1.0, and 0.3.

In magnetically driven simulations, we find \alfvenratio\ $\approx 0.8$, while in kinetically driven, this is \alfvenratio\ $\approx 2.5$. Both driving schemes suggest a linear scaling between magnetic and velocity fluctuations ($\delta u \sim \delta B$). When combined with the highly compressible simulations, the results indicate a Mach number-dependent transition from $\delta B \sim \delta u$ to $\delta u \sim \sqrt{B_0 \delta B}$. 

Magnetically driven simulations result in a balanced cascade that consists of locally imbalanced patches, consistent with the dynamic alignment theory. The turbulence cascade scales as $k^{-3/2}$. Alfv\'enic fluctuations drive the dynamics of these simulations. The properties of the joint distribution between the normalized cross-helicity and residual energy are consistent with solar wind turbulence; however, there is no net residual energy in the inertial range of these simulations. 

Kinetically driven simulations show significantly different properties. The cascade scalings of kinetic and magnetic power spectra are close to -1, likely due to the injected inhomogeneities, which can sustain the coupling between the Elsasser fields over a wide range of scales. There is a net and positive residual energy throughout the inertial range. The scaling of the residual energy (\Eresidual~$\propto k^{\alpha}$) changes with $\beta$. For $\beta = 4.0$, $-2 \lesssim \alpha \lesssim -5/3$, for $\beta = 1.0$, $-5/3 \lesssim \alpha \lesssim -3/2$, and for $\beta = 0.3$, $\alpha \approx -1$.

\bibliographystyle{aa}
\bibliography{bibliography}
 
\begin{acknowledgements}
We thank the anonymous reviewer for their helpful comments, which have significantly improved the clarity and quality of the manuscript. We thank J. Schober, and A. Brandenburg for useful feedback on the Pencil code. Many thanks to J. Squire, A. Polychronakis, and N. Soliman for fruitful discussions and to K. Tassis for providing useful comments on the manuscript. Support for this work was provided by NASA through the NASA Hubble Fellowship grant \#~HST-HF2-51566.001 awarded by the Space Telescope Science Institute, which is operated by the Association of Universities for Research in Astronomy, Inc., for NASA, under contract NAS5-26555. A.~Tritsis acknowledges support by the Ambizione grant no. PZ00P2\_202199 of the Swiss National Science Foundation (SNSF), and the MERAC Foundation. J.~R.~B. acknowledges funding from the Natural Sciences and Engineering Research Council of Canada (NSERC, funding reference number 568580) and support from NSF Award 2206756. We acknowledge support from NSF ACCESS \citep{nsf_access}. This work used Stampede 3 at Texas Advanced Computing Center (TACC) through allocation PHY240300 from the Advanced Cyberinfrastructure Coordination Ecosystem: Services \& Support (ACCESS) program, which is supported by U.S. National Science Foundation grants \#2138259, \#2138286, \#2138307, \#2137603, and \#2138296.
\end{acknowledgements}

\begin{appendix}

\section{Numerical convergence}
\label{sec:numerical_convergence}

\begin{figure}
   \centering
   \includegraphics[width=\hsize]{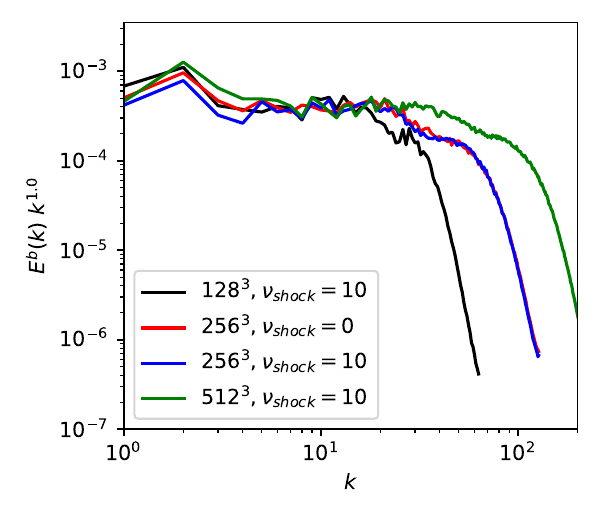}
   \caption{Kinetic power spectra of numerical simulations with different resolutions and bulk viscosities. The obtained power spectrum properties are consistent between the different runs, confirming the convergence in our results.}
   \label{fig:convergence_tests}
\end{figure}

Figure~\ref{fig:convergence_tests} visualizes kinetic power spectra of kinetically driven simulations with $\beta = 0.3$, different resolutions and $\nu_{shock}$ as indicated in the label. All spectra have been compensated by $k$. The scaling of the inertial range, $3 \leq k \leq 40$, does not depend either on the resolution or on $\nu_{shock}$. This guarantees the validity of our results against variations in these parameters. 

In $512^3$ simulations, the inertial range transitions to a steeper scaling at $k\sim 30$. A new inertial range with an IK scaling seems to emerge in the range $30 \leq k \leq 100$. This might indicate the transition of turbulence from large-scale weakly interacting to small-scale maximally interacting Alfv\'enic interactions, which is consistent with our interpretation about the cross-helicity variations of kinetically driven turbulence (Sect.~\ref{sec:theoretical_consideration_kinetic}). Higher resolution simulations are required to confidently establish the validity of this transition of the turbulent cascade. 

\end{appendix}

\end{document}